\documentclass[preprint,1p,twocolumn]{elsarticle}

\usepackage{lineno}
\usepackage{amssymb}
\usepackage{subcaption}
\usepackage{comment}
\usepackage{amsmath}
\usepackage{graphicx} 
\usepackage{xcolor}
\usepackage[symbol]{footmisc}
\usepackage{url}

\journal{Nuclear Instruments and Methods in Physics Research A}

\begin{document}


\title{Design and Performance Studies of a Granular Thin HCAL–MuID Detector for the EIC Optimized for AI-Based Reconstruction}




\author[duke,michigan]{Rowan Kelleher}
\author[duke]{Anselm Vossen}
\author[usc]{Yordanka Ilieva}
\affiliation[duke]{Duke University,Durham, North Carolina,27708,USA}\affiliation[michigan]{University of Michigan, Ann Arbor, Michigan, 48109, USA}
\affiliation[usc]{University of South Carolina,Columbia,29208,USA}
\author[iu]{William~W.~Jacobs}
\affiliation[iu]{Indiana University, CEEM, Bloomington, Indiana, 47508, USA},

\author[usc]{Pawel Nadel-Turonski}
\author[duke]{Simon Schneider}
\author[iu]{Gerard Visser}


\begin{abstract}
    We describe the design concept and estimated performance of an iron-scintillator sampling calorimeter for the future Electron Ion Collider. The novel aspect of this detector is a multi-dimensional readout coupled with foreseen excellent timing resolution, enabling time-of-flight capabilities as well as a more compact overall assembly. 
    Machine learning has been integrated into the detector design process from the ground up. Detector design objectives are defined using Machine Learning based reconstruction and Machine Learning is used to optimize the detector design. The highly segmented readout is 
    implemented with Machine Learning algorithms in mind to reach performance levels usually reserved for much more
    expensive detector systems.
    The primary physics objective 
    is to serve as a muon detector/ID system and a neutral hadron calorimeter.
    In EIC kinematics, charged particles are best measured through tracking rather than calorimetry, but the hKLM can identify and  measure the momentum of neutral hadrons.
    The latter are mainly $K_L$'s and neutrons: for lower energies, excellent relative momentum measurements of a few 10\% are achieved using time of flight, while for higher particle momenta, the energy can be measured calorimetrically with a resolution significantly better than that demonstrated for similar calorimeters 
    read out with less granularity. 
\end{abstract}
\maketitle
\section{Introduction}
Organic scintillators coupled with silicon photomultipliers (SiPMs) have emerged as a cost effective detector technology in particle physics. Recent applications 
include calorimetry~\cite{Liu:2025hgf, Laudrain:2025hnj, GlueX:2020idb, tsai2015development} and tracking~\cite{mazziotta2022light,pillera2023compact}.
Here, the scintillator materials are shaped as fibers, tiles, strips, or bars. Fibers have the advantage of good position resolution, whereas tiles (which also can  be read out by fibers on or near the 
periphery),
can provide a hit position
in two dimensions with a moderate channel count. 
A typical design for Time-of-Flight (ToF) uses scintillator bars 
coupled with photomultiplier tubes~\cite{smith1999time, CARMAN2020163629, gilman2017technicaldesignreportpaul}. In more recent designs,
the latter are being replaced by SiPMs~\cite{gruber2016barrel, Wang:2025stj}. 
Compared to
PMTs, SiPMs bring compactness (along with magnetic field immunity) and comparable or 
better performance at an increasingly low cost, making multiple sensor readout feasible.
Resolutions of better than 50--100 ps, depending on scintillator length, can be routinely reached using these techniques~\cite{smith1999time, gruber2016barrel, Wang:2025stj}. At the same time, more cost-effective readout electronics as well as 
application
of advanced analysis tools, \textit{e.g.}, utilizing Machine Learning (ML), enables the incorporation of additional information available from a finer longitudinal and lateral readout segmentation along the detected particle's path.

Our contribution
is to combine these aspects in a novel way to design a low-cost, compact, high-performance detector simultaneously providing muon identification, hadronic calorimetry, and time-of-flight determination
for the EIC. 
Compared to some recent tile-based iron-scintillator sampling calorimeters~\cite{CALICE:2024jke}, this detector provides better timing resolution with less complexity while
maintaining sufficient granularity for the event multiplicities expected at the EIC. In comparison with an implementation using scintillator
strips read out with Wave-Length-Shifting (WLS) fibers, such a detector has
better timing resolution and
can in principle be more compact
since no orthogonal layers are needed.
A key aspect of this work is to 
demonstrate how to use ML/AI in an integrated way for the design, evaluation, and simulation of the detector and, in addition, carry out systematic multi-objective performance optimization and evaluation. 
We find that one can surpass the state-of-the-art performance for each targeted capability. For hadronic energy resolution, the results 
indicate that one is competitive with any HCAL, regardless of the HCAL technologies currently deployed. 

The proposed combined hadronic calorimeter (HCAL) and muon identification (MuID) detector uses fast timing with a 
cost-effective design utilizing multiple SiPMs reading out thin scintillator bars (strips) on both ends. Separate 
longitudinal (along the radius of the calorimeter) and lateral (along the strip) information, combined with the use of Machine Learning for energy reconstruction, enables a better performance of the calorimeter than a more coarse segmentation. Fast timing from the scintillators provides time-of-flight information for particles 
as well as the localization of hits along the 
strip.
The detector must simultaneously fulfill several objectives that pose competing constraints on the design parameters, such as the thicknesses of scintillator and passive materials. Performance goals include: good energy and timing resolution as well as muon identification with a low fake rate. We therefore use a multi-objective optimization in the detector design as discussed further below. In line with the naming of the Belle $K_L$ and muon detection system, KLM~\cite{Wang:2003bm}, we call the concept discussed here hKLM to emphasize the neutral hadron detection capabilities. 
The principles and insights discussed here are general and thus have broader application. For the study presented here, however, we consider a specific design for the barrel region of a potential $2^{\text{nd}}$ detector at the EIC~\cite{Gamage:2022rzd}.

\section{Overview and Physics}
The hKLM is designed
to serve as a thin HCAL and MuID detector integrated in the flux return steel of a potential $2^{nd}$ detector at the Electron Ion Collider. At the top level,
the design consists of scintillator strips read out by SiPMs on both ends, alternating with the steel of the flux return. This layout can also be found in the Belle II KLM and in the proposed CORE detector for the EIC~\cite{CORE:2022rso}. However, in these instances timing and finely segmented readout were either not implemented or were not studied in detail.
A rendering of the barrel region with iron-scintillator layers for a KLM-type detector for the EIC can be seen in Fig.~\ref{fig:KLMDesign}. The physics objectives for the hKLM MuID capabilities include the identification of muons down to below 1 GeV/$c$. This value is driven by reaction channels of interest at the EIC, such as quarkonium production~\cite{AbdulKhalek:2021gbh} and timelike Compton processes like double Deeply-Virtual-Compton-Scattering (DVCS)~\cite{Deja:2023lqc}. The hKLM HCAL capability will also play an important role in the jet physics program at the EIC~\cite{AbdulKhalek:2021gbh}. Multiplicities will be low, with the majority of jets having less than ten particles. The task of the HCAL is, therefore, mainly to identify and measure neutral particles (neutrons and $K_L$) that are present
in about a third of the jets~\cite{AbdulKhalek:2021gbh}.
The hadron energy can be measured either using the ToF at lower momenta, or with a calorimetric measurement at higher momenta. Section~\ref{sec:performance} discusses the performance in the two different 
ranges in more detail.

\section{Design}
\label{sec:Design}
The starting point of the design is the KLM of the CORE detector proposal~\cite{CORE:2022rso}, itself derived from the Belle KLM and adapted to the boundary conditions of the interaction region (IR) 8 at the EIC, the likely home of the second detector. The barrel design is envisioned as an octagonal structure surrounding the beam pipe. At the inner radius of $r=1770$~mm, each side of the octagon is a rectangle with length $l_z=1500$~mm along the $z$ axis and width $l_x= 1465$~mm in the perpendicular ($\phi$) direction. These rectangles form the base of the different projective layers in each sector of the detector. Each sector has 14 layers of alternating iron and scintillator layers each, with the outer radius constraint given by the IR configuration to be $r_{out}=2835$ mm. 

\begin{figure}
\includegraphics[width=0.49\textwidth]{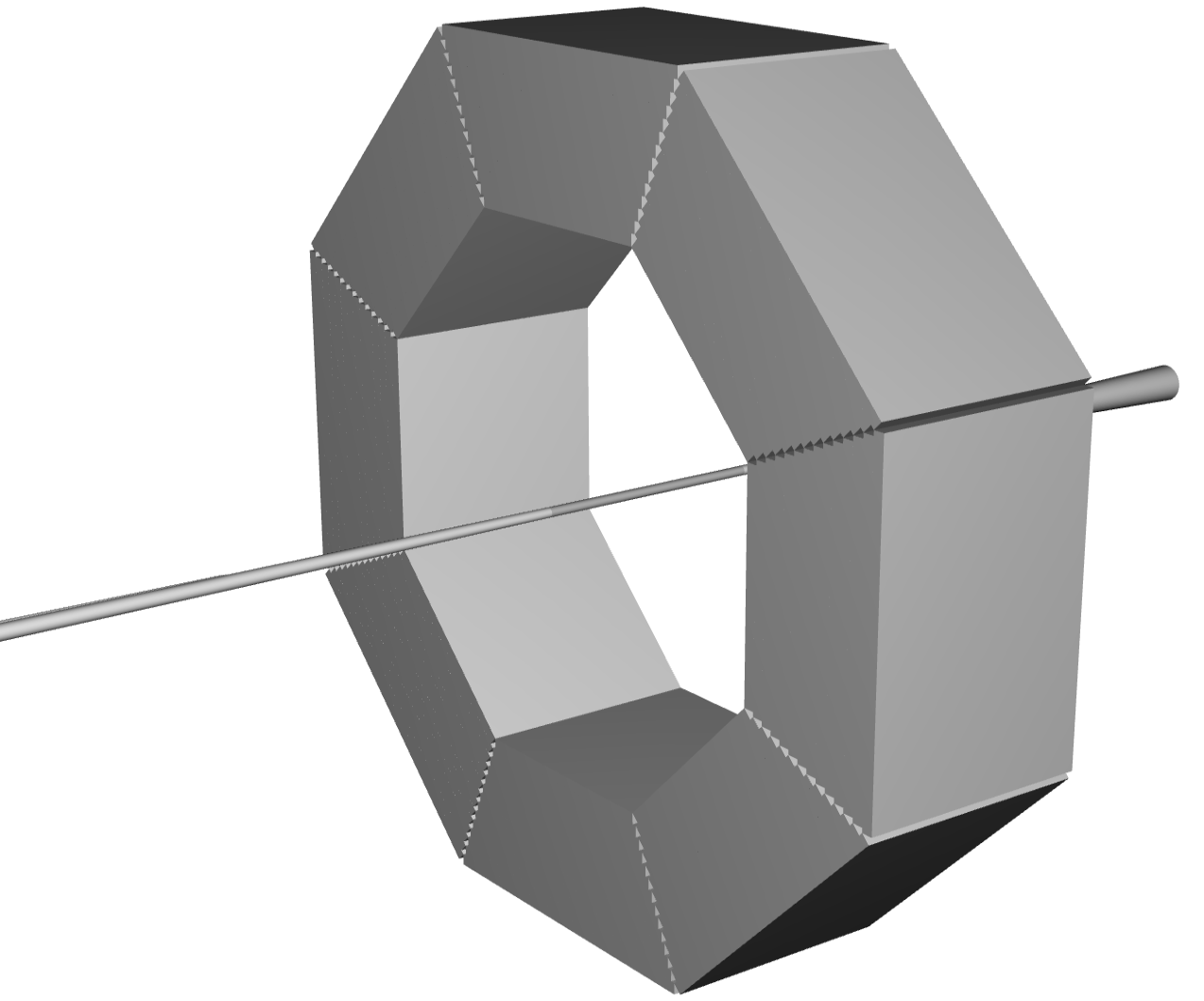}
\includegraphics[width=0.49\textwidth]{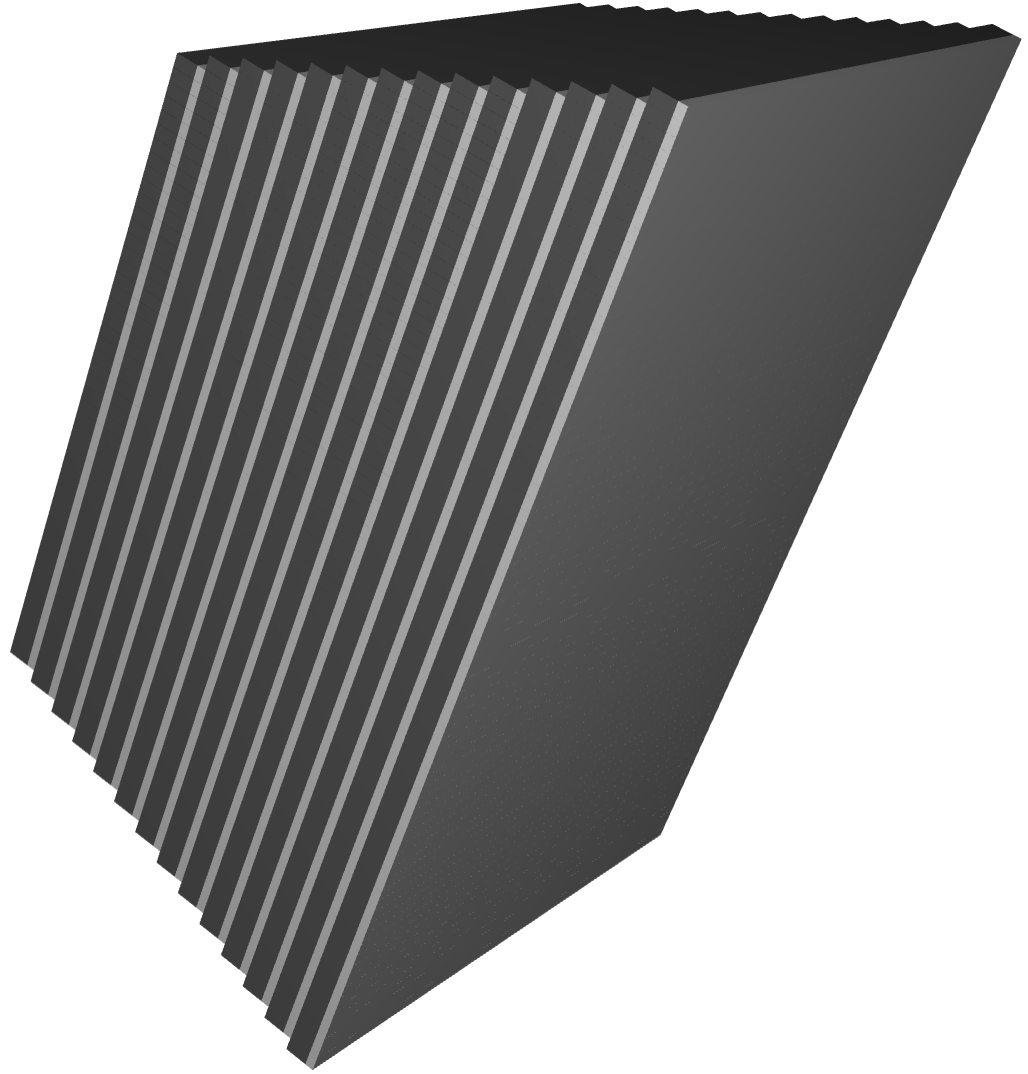}
\caption{\label{fig:KLMDesign}Octagonal arrangement of sectors around the beampipe (left) and iron/scintillator sandwich structure of one sector (right).
}
\end{figure}

This design is shown in Fig.~\ref{fig:KLMDesign}.
The relative thicknesses of the active and passive layers are subject to optimization as is outlined in Sec.~\ref{sec:optimization}. Depending on the objective, the optimal configuration varies. MuID at large muon momenta requires more absorption length, while energy reconstruction becomes more precise with thicker active layers, as long as the shower is contained in the hKLM. There is a minimum amount of steel that has to be available in the flux return in order to contain the full magnetic field, which in turn depends on the chosen magnet strength and geometry, as well as constraints on the fringe fields around the detector. A reasonable assumption for a 1.5-T magnet in our case is a minimum of at least $\sim$ 50\% steel by thickness in the hKLM. For comparison, in the CORE design for a 3~T solenoid, which followed as a more compact
version of the Belle KLM 
(which in turn was optimized for muon detection), an insertion gap of 21.5 mm and 55.5 mm thick steel plates was used. This corresponds to about 72\% steel. The status quo design uses these thicknesses with 14 layers, giving 4.87 nuclear interaction lengths from the inner radius to the outer radius. With recent updates to the EIC layout eliminating the Rapid Cycling Synchrotron (RCS) control line which was previously located close to the interaction point, the higher iron fraction in a stronger solenoid can be traded for a slightly larger outer radius, opening up another avenue for optimization.

The active layers of the hKLM consist of scintillator strips  positioned along the $z$ (beam-axis) direction and are directly read out with SiPMs on both sides. 
This is at variance to the Belle II KLM, 
with a single SiPM reading out one end of a WLS fiber running in the cross-sectional middle along the length of the strips, with two planes of strips per layer in order to reconstruct a 2D point. Here we use timing from the readout located at the two ends to locate the hit along the strip. We plan to investigate, using pulse shape discrimination techniques, the possible resolution of double hits in order to provide information on the shower shape. The multiplicity and interaction rates at the EIC are moderate so that the probability of multiple primary particles incident on the same bar is low. 
While the thickness of the active layers is subject to optimization, the width of the strips is taken to be 3~cm. 
In order to account for the effect of SiPM coverage (packing fraction) on the strip ends, we assume from an initial strip thickness and chosen SiPM array, that this fraction stays constant as we vary the thickness during optimization studies.
\section{SiPMs and Readout}
\label{sec:readout}
For the purpose of this study we assume scintillators using EJ-204 and Hamamatsu S14160-4050HS photo sensors. This scintillator combines fast timing and a high scintillation efficiency with an attenuation length of 1.6~m, which makes it suitable for the strip length of 1.5~m studied here.

 These SiPMs have a peak sensitivity wavelength that is well matched to the scintillating material. The photo detection efficiency is about 50\% at the peak sensitivity wavelength. Their effective photosensitive area of 4×4 mm$^2$ allows us to mount a cluster of them on
a readout board in order to cover a large fraction of the nominal 1×3 ~cm$^2$ cross section of the scintillator strip end. The readout board developed as part in this work carries 12 SiPMs, which cover 64\% of a bar’s cross section, and is based on the design used in the HELIX experiment~\cite{Park:2021oic}.
In the present application, for faster pulse and lower electronics noise,
we combine three SiPMs in
parallel into one load resistor and timing preamplifier, instead of all
twelve in parallel. The outputs of these four timing preamplifiers are then
summed in a second stage amplifier and cable driver.
The simulation uses the rise and fall times measured for this circuit, which are 1.1~ns and 15~ns, respectively.
(Note: for the signal amplitude, a slow current pulse is tapped off each of the four groups
of three SiPM's in parallel 
and is summed and amplified to drive another signal cable.)
Biasing on each of the 12 SiPMs can be individually adjusted for testing
flexibility and/or output alignment purposes.

The amplified timing signal can be read out with various digitizers that have sufficient timing resolution, such as commercially available~\cite{Bitossi:2016waj,petiroc,pillera2025beam,triroc,ahmad2015triroc} or electronics under development ~\cite{temporoc}.
For the simulated timing measurements of this study, we use a leading edge discriminator; for the energy measurements we additionally 
simulate an integrated signal output.

\section{ML/AI assisted Simulations}


The hKLM design is implemented in the Detector Design for High Energy Physics framework (DD4HEP)~\cite{frank_markus_2018_1464634}. The hKLM geometry is built from a compact XML file, and interfaces with the GEANT4 simulation software via DD4HEP. The framework provides a python based interface with GEANT4, where input files from event generators (hepmc3 or ROOT file types) or particle gun definitions can be used to specify particle generation. All simulations use GEANT4 version geant4-11-02-patch-01 with the physics list ``FTFP\_BERT''. The ``Geant4CerenkovPhysics'', ``Geant4OpticalPhotonPhysics'', and ``Geant4ScintillationPhysics'' plugins are loaded when simulating optical photons, but are not used when utilizing the fast parametrized generator described in this section.

As discussed above, the detector is a steel-scintillator sandwich design. Thin air gaps (0.3 mm) are inserted between each material layer. At variance to the earlier designs, we employ only a single scintillator layer per steel layer with strips positioned (without orientation prejudice) along the $z-$direction
with the hit location along the strip determined by the signal arrival timing at the two ends. We use a strip width of 3~cm and a strip length of 150~cm. The number of layers and thickness of steel and scintillator layers are subject to an optimization described below.
The starting point for the design is 14 layers with a steel to scintillator split of 72\% to 28\%. We will refer to this geometry as the status quo (SQ).


As the detector optimization process explores a large number of design parameters, the resources required for the simulation are significant. The leading contribution is the simulation of the optical photons. We therefore decided to replace it with a fast parametrized generator utilizing generative Artificial Intelligence (AI). This approach proved to fulfill both of our requirements, a significant processing speedup while still faithfully reproducing the distribution of the incoming photons, which is an important input to downstream particle ID and the energy reconstruction algorithms.


The parametrized approach sped up the simulation sufficiently to enable automatic optimization, which needs to run a significant number of events for each evaluation point of the design. The speedup factor was roughly 20.
    Our specific implementation uses a parameterization for the mean number of photons emitted for a given track energy and angle. For each photon that is sampled from a distribution with the given mean, the arrival time and position are determined using a normalizing flow (NF)~\cite{rezende2016variationalinferencenormalizingflows} model that uses the emission point of the photon. Normalizing flow is an appropriate model to use here due to the modest dimensionality of the parameter space and since it enables us to sample efficiently from the learned distribution. 

For the normalizing flow 
we use a conditional 
model, implemented with normflows~\cite{Stimper2023} with 12 flows using autoregressive rational neural spline flows~\cite{durkan2019neuralsplineflows}. The total model has about 41 million parameters.
The NF model is conditioned on the impinging charged track momentum, angle, and its distance to the SiPM, 
the previously sampled number of photons as well as the energy deposited in the scintillator by the track.
We observed that a model with fewer flows (and thus fewer parameters) was able to give a quite reasonable description of the overall photon distribution. However, to accurately describe the timing distributions of the ordered photons (1$^{\textrm{st}}$, 2$^{\textrm{nd}}$, \textit{etc.}), we chose to use 12 flows for the simulations described in this study.

\begin{figure}
\centering
\includegraphics[width=0.49\textwidth]{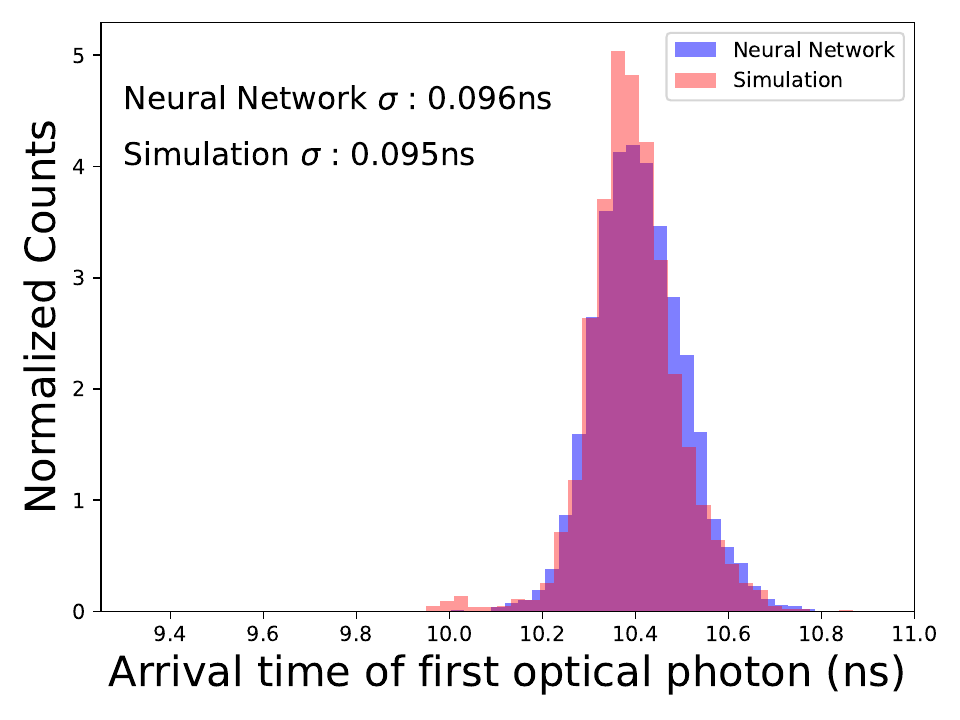}
\caption{Comparison of arrival time distributions of the first photon in the full simulation and the parametrization with normalizing flow.\label{fig:NN_vsSimu}}
\end{figure}

The combined model, parametrization plus normalizing flow, reproduces the full optical photon simulation very well. This is shown in Fig.~\ref{fig:NN_vsSimu}. We also validated our optical photon simulation with timing resolution measurements for 
specific
scintillator geometries in the literature~\cite{gruber2016barrel, Wang:2025stj,Bohm:2016jhc}. 

For the ML assisted performance optimization, we trained a NF for two scintillator thicknesses. We confirmed that the differences between the models for different thickness under consideration were small. During optimization, we picked the NF model that was trained with a scintillator thickness closest to the configuration to be evaluated.

\section{Performance}
\label{sec:performance}
This section describes the performance of the detector setup in the simulation framework described above. For each simulation we generate particles with momenta between 0.5~GeV/$c$ and 5~GeV/$c$. We refer to the 0.5~GeV/$c$ to 2.75~GeV/$c$ range as the ``low energy'' range, and from 2.75~GeV/$c$ to 5~GeV/$c$ as the ``high energy'' range. We then specifically study the performance for the following tasks.
\begin{itemize}
\item MuID in the low and high energy ranges.
\item Energy reconstruction for neutrons and $K_L$, the particles relevant at the EIC, in the low and high energy ranges.
\item Determination of time of flight for hadrons that can be used to determine energies in conjunction with the PID.
\end{itemize}

\subsection{Muon Identification Performance}
\label{sec:MuID}

\begin{figure}
\centering
\includegraphics[width=0.8\textwidth]{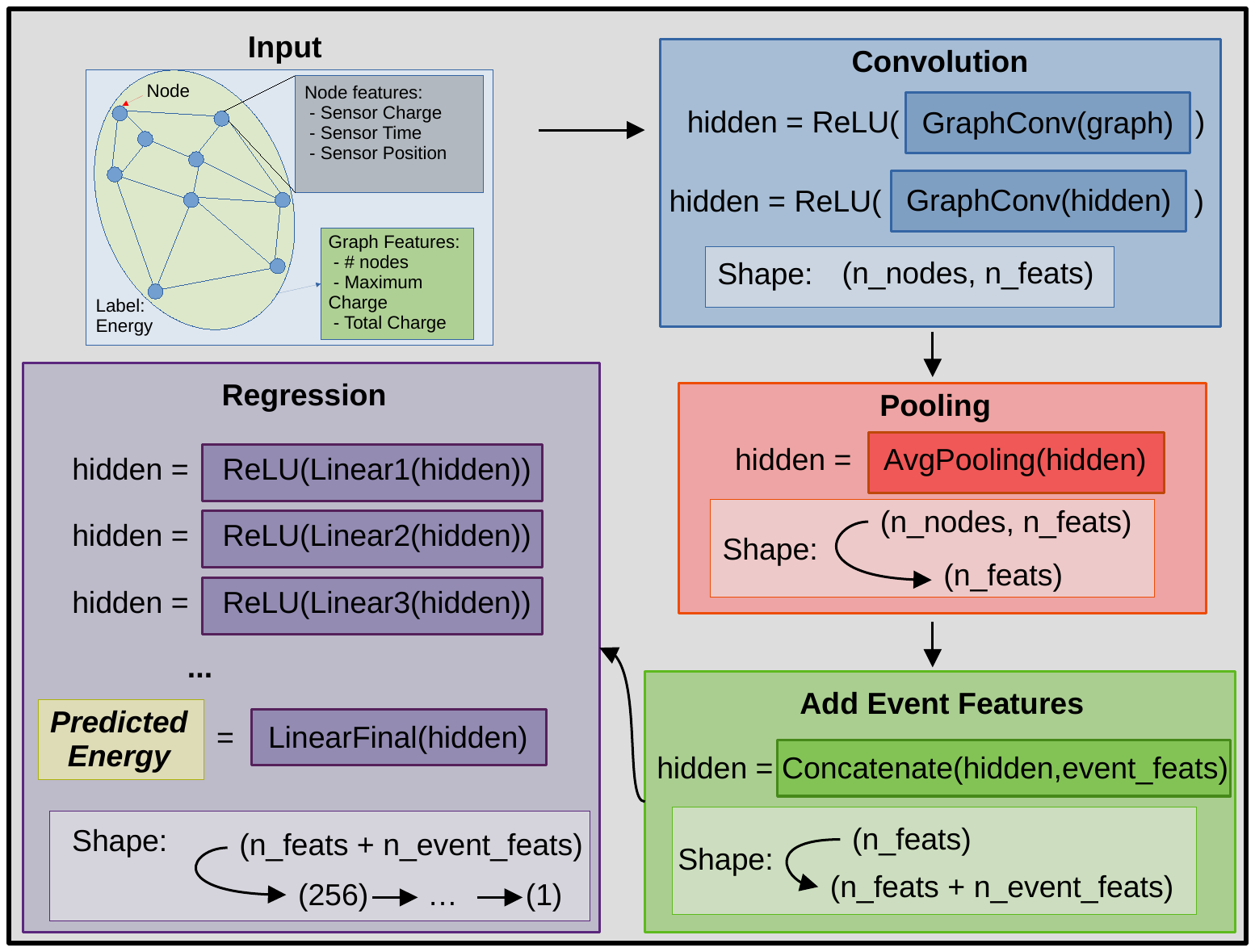}\\\includegraphics[width=0.5\textwidth]{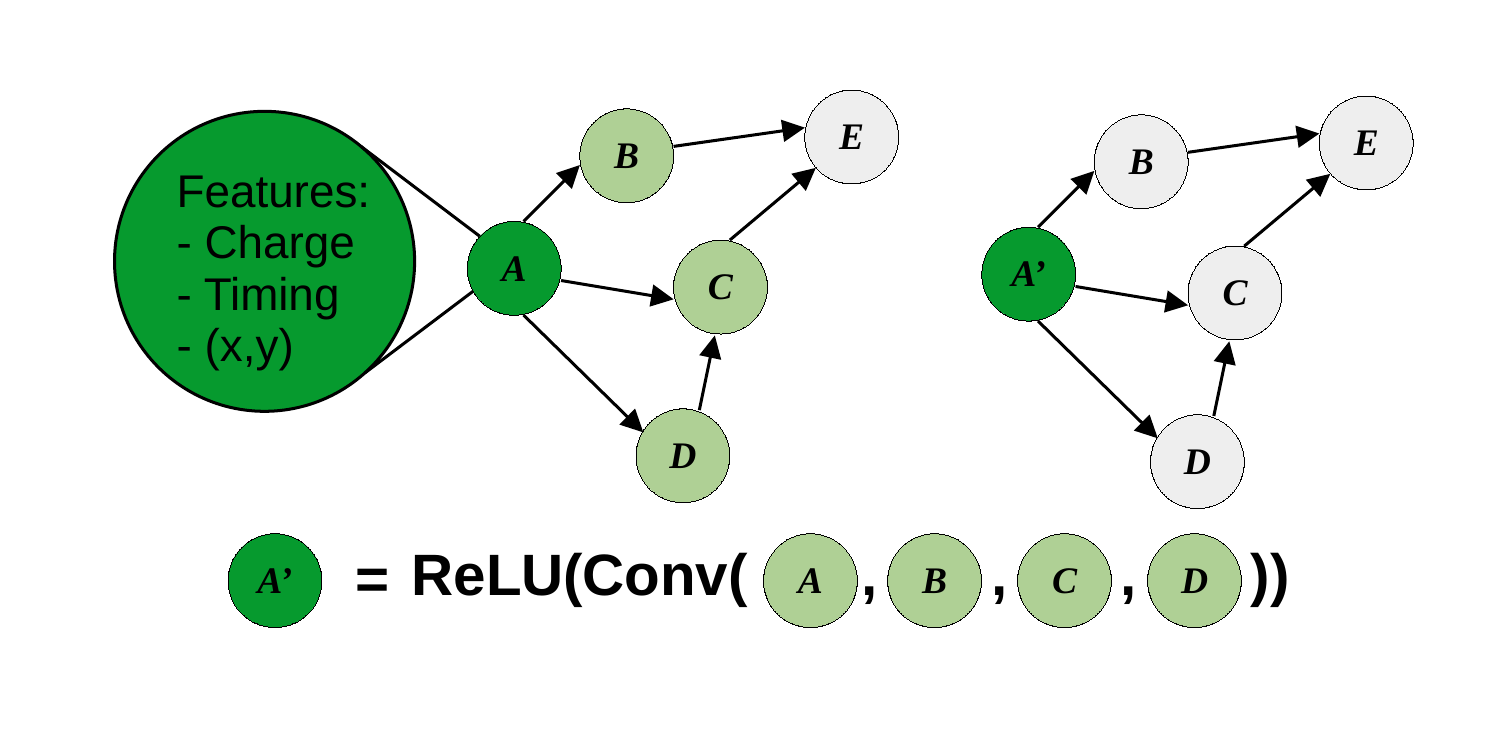}\\
\includegraphics[width=0.9\textwidth]{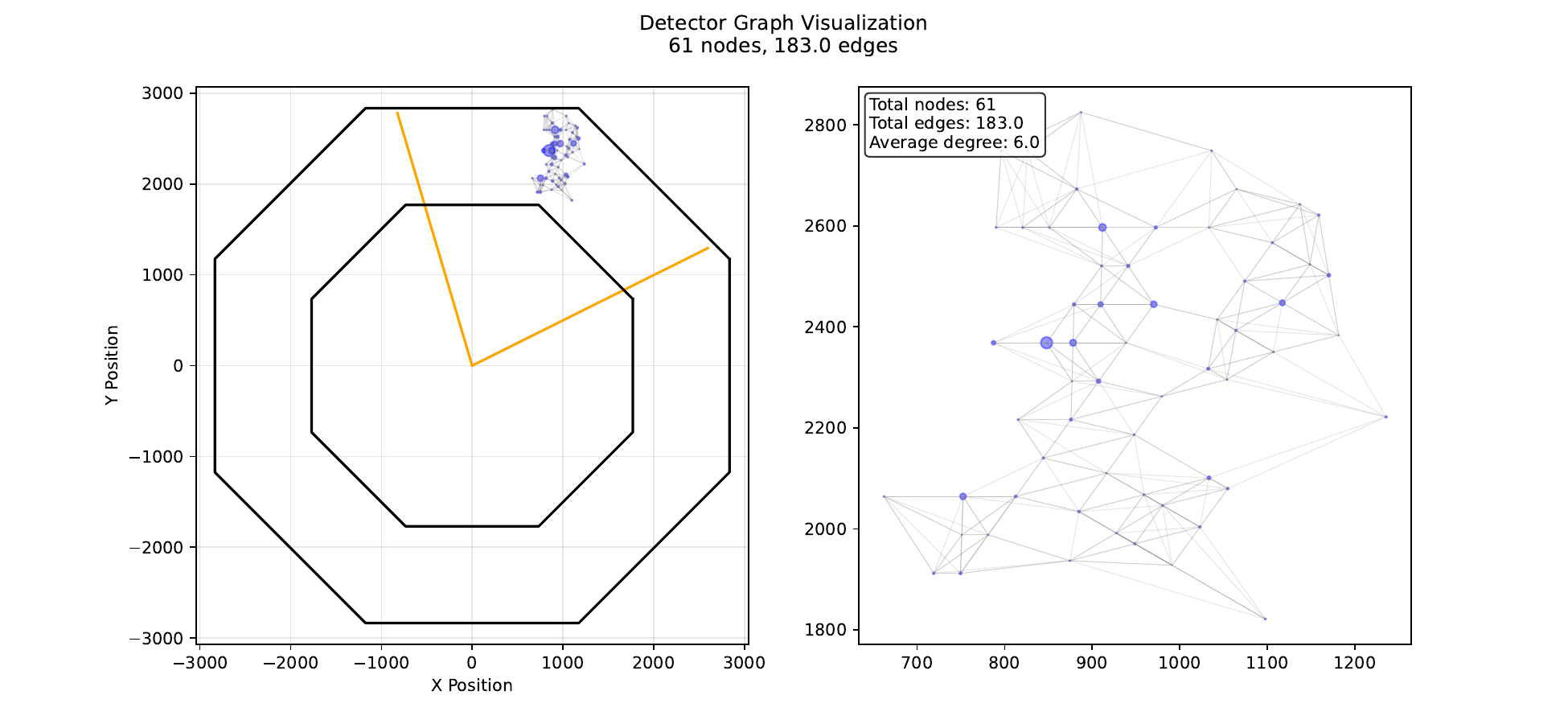}
\caption{Top: GNN Architecture including input and graph features used by the GNN. Middle: Graph Operation, Bottom: Example of the graph structure used by the GNN in  a specific event. The graph has edges between the five nearest hits in the detector.\label{fig:GNN}}
\end{figure}

We study the performance of a Graph Neural Network (GNN) for MuID. The segmentation in the radial and azimuthal directions provides shower shape information which enables the GNN to exceed the performance of a conventional approach. The GNN architecture is show in Fig~\ref{fig:GNN}. The GNN uses the sensor charge, signal time, and scintillator position as node features, and the number of nodes, as well as maximum and total charge, as graph features. For each particle species, 10,000 events are simulated. The dataset is split into three categories: 7,000 events are used for training the GNN; 1,500 events are used for validation throughout the training process; and the final 1,500 events are used for testing the GNN.

\begin{figure}
\includegraphics[width=0.49\textwidth]{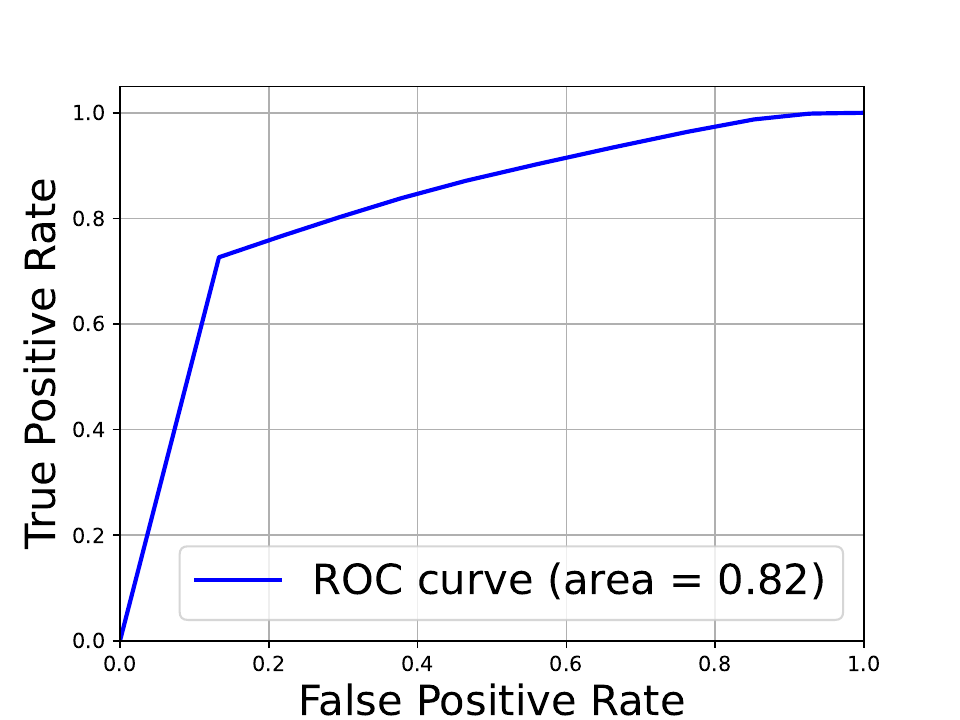}\includegraphics[width=0.49\textwidth]{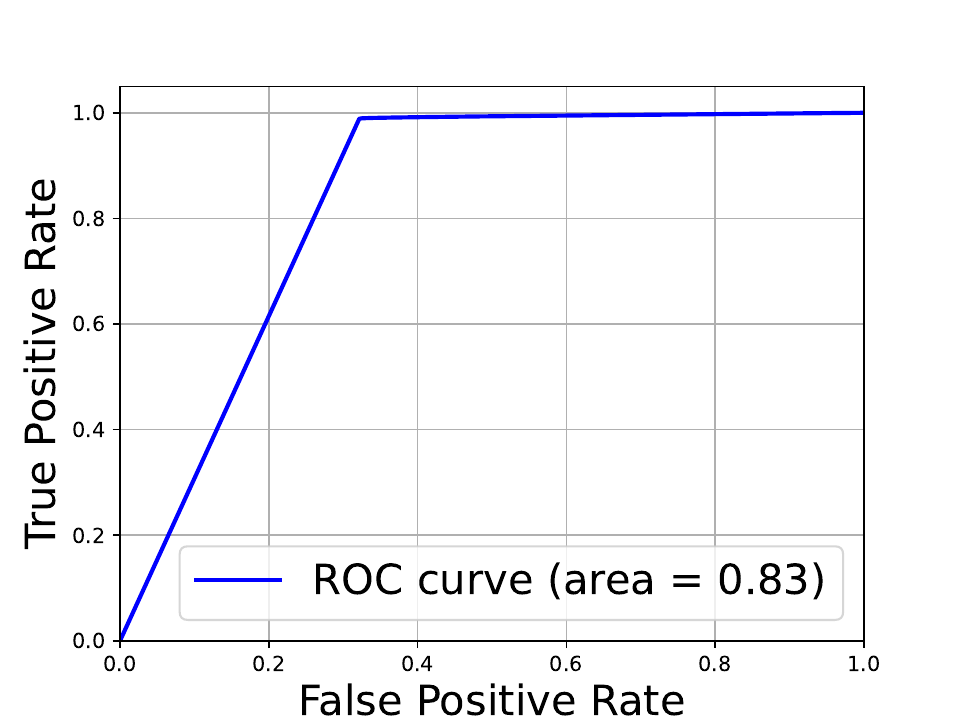}
\includegraphics[width=0.49\textwidth]{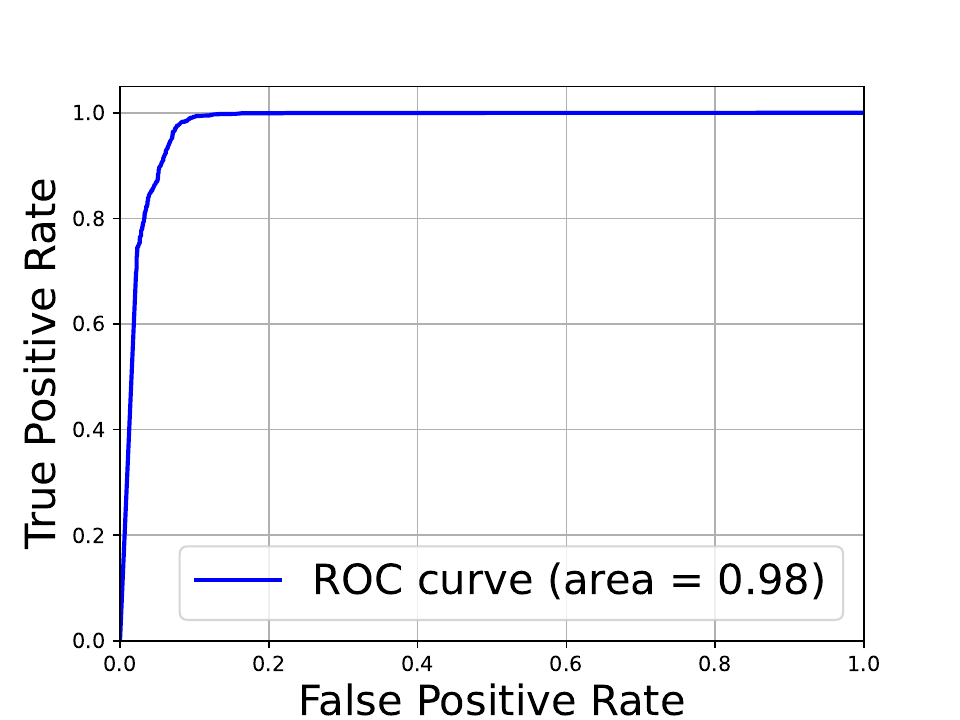}\includegraphics[width=0.49\textwidth]{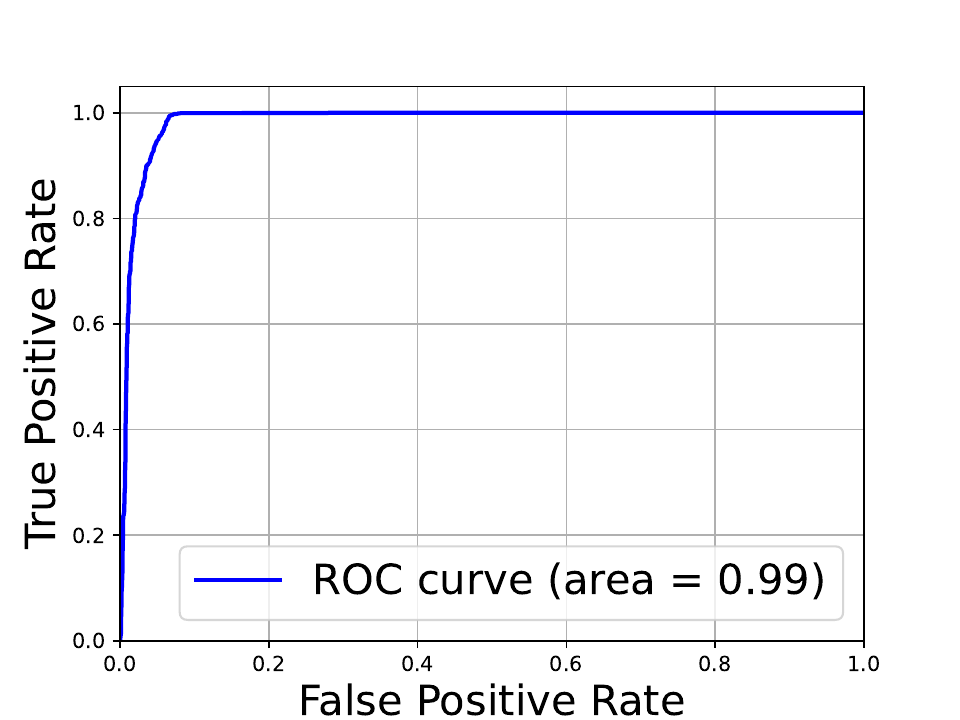}
\caption{\label{fig:muIDPerformance}ROC curves for MuID performance with the status quo design. Plots at the top use the conventional ID method for muons with momenta between 0.5~GeV/$c$ and 2.75~GeV/$c$ (left) and between 2.75~GeV/$c$ and 5~GeV/$c$ (right). The bottom plots show the performance using a GNN for muons with momenta between 0.5~GeV/$c$ and 2.75~GeV/$c$ (left) and between 2.75~GeV/$c$ and 5~GeV/$c$ (right).}
\end{figure}

We compare results using the general performance criteria for a classification model as deduced from the ROC (Receiver Operation Characteristic) curve relating the positive rate \textit{vs.} the false positive rate. The area under the ROC curve (AUC) is a performance measure independent of a specific desired purity/efficiency.
Using this ML method one finds an impressive MuID performance, with a 0.99 area under the roc curve in simulation, as shown in Fig.~\ref{fig:muIDPerformance} on the right.
For comparison with the ML approach, a basic MuID algorithm was implemented in analogy to the one used at Belle II. This approach is
based on the measured momentum, and the number of layers a pion and a muon are expected to traverse is compared with the actual penetration depth.
Figure~\ref{fig:muIDPerformance} (two leftmost plots) shows the performance of the MuID which, as expected, is comparable to what was achieved at Belle with a similar design~\cite{Wang:2003bm}. 
Note that the true positive rate for the muons with momentum between 2.75~GeV/$c$ and 5~GeV/$c$ jumps from 0 to around 0.95 because nearly every muon reaches the final layer; the conventional method predicts that all of these are either muons or not muons, resulting in an all-or-nothing result. For both momenta ranges, the ML method performs better than the conventional method. The difference is most pronounced for high-momentum tracks where the conventional method suffers due to punch-through of the pions.

\subsection{Hadronic Energy Measurements}
\label{sec:EMeasurement}

Compared with previous approaches which only use the summed response of the scintillator material~\cite{ALEPH:1994ayc, DELPHIHAC:1996wto}, energy measurements can be significantly improved by exploiting 
finer segmentation. Here we use segmentation in the radial direction (layers) of the readout of the barrel detector configuration as well as perpendicular to the radial direction by considering the signal from each scintillator sensor strip 
as well as the respective photon timing information. These quantities can also give access to some shower shape information. To make use of the more complex correlations in these data, we utilize a GNN with a structure similar to that for the MuID measurements described in Sect.~\ref{sec:MuID}. The GNN is trained using the same train-validation-test split of 10,000 events per geometry.

With the current electronics design, the timing resolution for a leading edge measurement is precise enough to determine the origin of a shower within a few cm, about the resolution of the strip width. 
In principle, electronics is available that can provide multi-GHz bandwidth
~\cite{aardvarc}, which might provide sensitivity to the inner shower structure (shower shape). This can be a possible avenue to investigate in the future.

\begin{figure}
\includegraphics[width=0.95\textwidth]{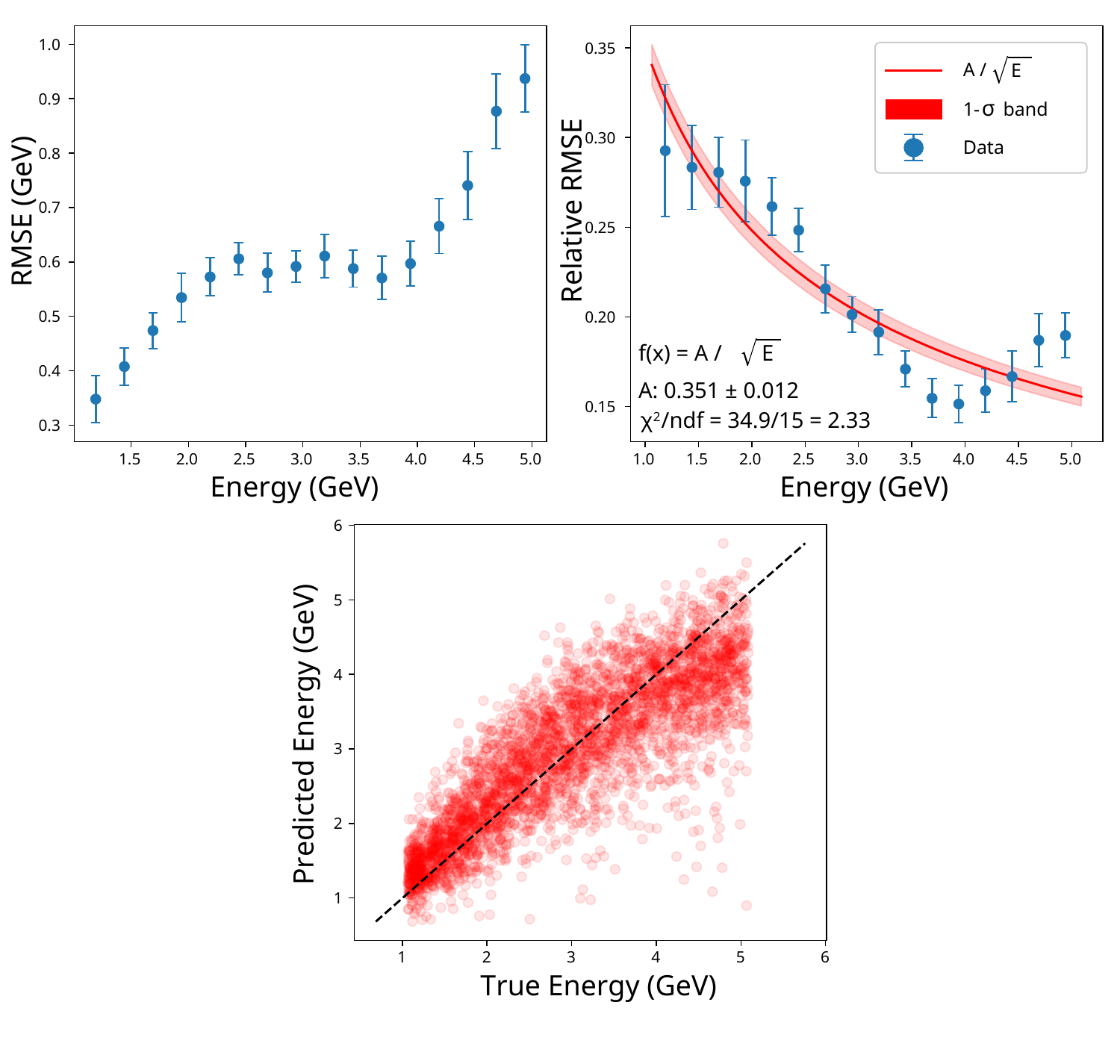}
\caption{\label{fig:E_resolution} Energy resolution for neutrons for the status quo design. Top left: total error. Top right: relative error. A fit (red curve plus $1-\sigma$ band) to the functional form $A/\sqrt{E}$ indicating a relative error of 35.1$\pm$1.2\%/$\sqrt{E}$. The goodness of the fit is evaluated using $\chi^2/\text{ndof} = 2.33$, where $\text{ndof} = 16 - 1$ for the 16 energy bins and 1 fit parameter. Bottom: true and predicted particle energy for neutrons in the test dataset (red scatter points), with y = x target line (blue), from a single GNN.}
\end{figure}

Figure~\ref{fig:E_resolution} shows the energy reconstruction performance for neutrons achieved with the current design. The leftmost plot shows the total error, the middle plot the relative error and the right plot the predicted vs. true energy particle energy. 
We estimate the total error using an ensemble of 30 GNNs. For each GNN, we randomize the training and testing subsets, train the GNN, and compute the RMSE per energy bin. The total error per bin is averaged across the models, and the standard deviation is reported in the error bars. 
Due to the nature of the simulation, we expect only the stochastic term to contribute to the resolution. As shown in the middle plot, the relative energy resolution follows the expected $1/\sqrt{E}$ dependence reasonably well, and the fit has a $\chi^2/\text{ndf} = 2.33$. For high energy neutrons, the hadronic shower can occasionally leak, causing the GNN to have difficulty reconstructing the energy. In tests we found that adding extra layers that help contain the complete shower help lower the RMSE above 4 GeV. The contour plots in Fig.~\ref{fig:moboBasic} support this conclusion, as the RMSE improves with more layers.

Even though the overall $\chi^2/\text{ndf}$ value is reasonable, a visual inspection shows disagreement in the high and low energy regions. As discussed, we attribute this to punch-through effects at high energies and large fluctuations in energy deposit at low energies. We expect the $1/\sqrt{E}$ relation to apply best in the region that suffers from neither the low energy fluctuations nor the high energy punch through. To demonstrate the stability of the result, we also investigated a fit to the energy range 2~GeV/$c$ $<$ $E$ $<$ 3.5~GeV/$c$. The restricted fit result $A = 0.356 \pm 0.008$ is consistent with the result of the fit to the full range, showing that the latter fitting method produces a reliable result.



The achieved relative energy resolution of (35.1 $\pm$ 1.2)\%/$\sqrt{E}$ is a significant improvement compared to sampling calorimeters using a similar sandwich design but less longitudinal segmentation, see {\it e.g.} Ref.~\cite{Bagliesi:1989eu}. While this is very encouraging, we note that there are still some limitations to the simulation, most notably the absence of realistic electronic noise, and potentially
other contributions.

To justify the use of the machine learning approach, we compare the energy reconstruction performance of the GNN to a conventional method. We estimate the particle energy with a linear regression using the total sensor charge:
\begin{equation}
    E = \sum_l A_l \cdot Q_{l,\text{tot.}} + B_l,
\end{equation}
where the sum runs over each layer $l$, and $Q_{l}$ is the sensor charge in layer $l$. Note that the GNN trained for this study and for Fig.~\ref{fig:E_resolution} are different, and differ slightly in their results due to statistical fluctuations and the stochastic nature of the ML model.

The ML and conventional energy reconstruction results are compared in Fig.~\ref{fig:GNN_vs_conventional}. The conventional method predicts a small band of energies around the mean for nearly all neutrons, whereas the GNN predictions follow the true energy more reliably. This result suggests that the fluctuations in the neutron shower from event to event are too large for a conventional method to successfully reconstruct the energy.
On the other hand, the GNN has a higher capacity to capture these fluctuations and the complex shower shapes to predict the energy very well. 


\begin{figure}
    \centering
    \includegraphics[width=0.95\linewidth]{neutron_GNN_vs_conventional_larger_labels_4x4.pdf}
    \caption{A conventional energy measurement method for neutrons (orange) compared with the GNN method as described by Fig.~\ref{fig:GNN} (blue). }
    \label{fig:GNN_vs_conventional}
\end{figure}

\subsection{Timing Performance}
\label{sec:timingPerformance}
\begin{figure}
\centering
\includegraphics[width=0.8\textwidth]{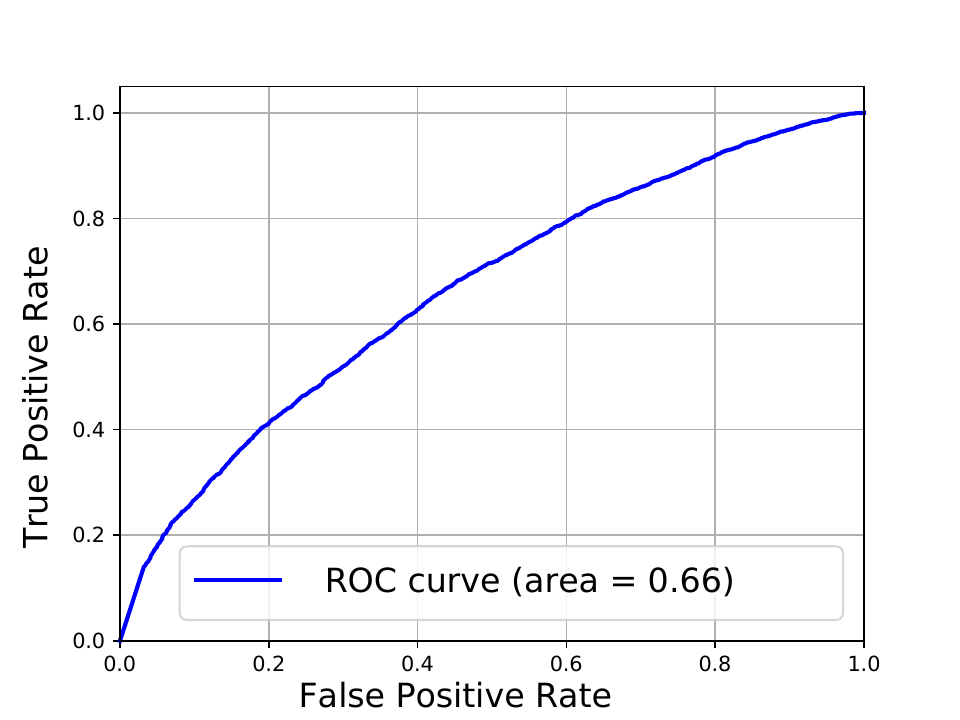}
\caption{ROC curve for differentiating neutrons from $K_L$ using a GNN with an analogue architecture as previously discussed for the energy reconstruction. Results for a sample with an equal number of neutrons and $K_L$ are displayed; positive examples refer to the correct identification of neutrons. Purities are modest, with an area under the curve of 66\%. Focusing on the lower left part of the ROC curve, purities of about 75\% can be reached at the cost of a lower efficiency.
\label{fig:pidROC}}
\end{figure}

\begin{figure}
\centering
\includegraphics[width=0.95\textwidth]{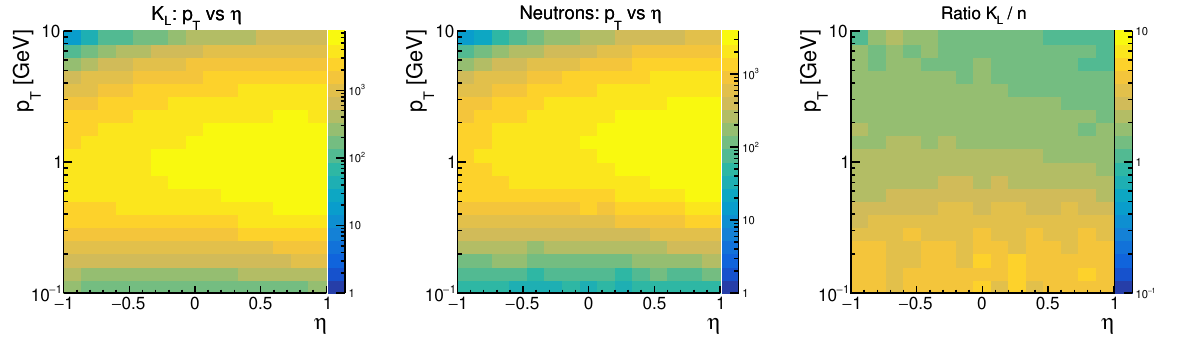}
\\
\includegraphics[width=0.95\textwidth]
{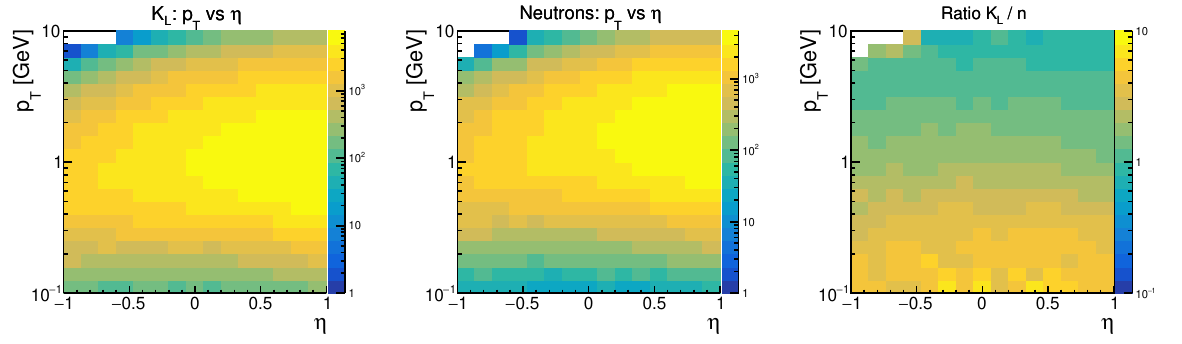}
\caption{Distribution of the transverse momenta of $K_L$ (left), neutrons (middle), and their ratio (right) $vs.$ pseudorapidity $\eta$ for SIDIS events with the condition $Q^2>1$ at the EIC. Only the barrel region $-1 < \eta< 1 $ currently instrumented by the hKLM is shown.
The top row shows the configuration with a 10 GeV/$c$ electron beam and 100 GeV/$c$ proton beam, the bottom row the configuration 18 GeV/$c$ on 275 GeV/$c$. 
Over most of the phase space, $K_L$ dominate neutrons by a factor of two to five.\label{fig:KLnRatio}}
\end{figure}

Good timing performance of the scintillators is a driver for the design of the scintillator and its readout. 
The goal is a timing resolution around 100~ps. This timing resolution would help with two different objectives. First, it would enable the determination of the position of a shower on a scintillator strip with a precision of about 2 cm. This would be similar to the strip width and would therefore enable a simplification of the design from two layers of orthogonal strips to a single layer in each steel gap. The other objective is to enable time-of-flight measurements. If one assumes a more conservative timing resolution of 150~ps, to include contributions for the start time determination, and a flight path of 1.8~m corresponding to the inner radius of the hKLM in the status quo design, a momentum resolution of about 15\% can be reached for $K_L$ mesons with a momentum of 1.2 GeV/$c$.
The same resolution can be reached for neutrons at a momentum of 2.2 GeV/$c$.
However, extracting a neutral particle momentum (and its energy) from the measured
time of flight 
requires particle identification (PID). Use of a GNN similar to that discussed earlier for the energy measurements
to extract a PID for neutral hadrons from the shower shape leads to only relatively modest purities for neutrons as shown in Fig.~\ref{fig:pidROC}. This is an issue that we plan to address in future studies. One complication might be that differential sensing of the shower shape along
a strip is not possible with the current setup that records only the integrated charge. We plan to study detector configurations with orthogonal strip orientations in different gaps and configurations where some gaps are instrumented with orthogonal scintillator strips. We are also performing R\&D to use waveform sampling to discern signals from close hits on the same scintillator. As shown in Fig.~\ref{fig:KLnRatio}, the a priori likelihood of finding a $K_L$ over a neutron is already larger by a factor of two to five. Therefore, a ToF measurement with a modest PID will already be helpful to constrain \textit{e.g.}, measured jet energies. As events containing neutrons are typically different from those containing $K_L$'s due to baryon number and strangeness conservation, we also expect that foundation models, taking into account the response of multiple detector systems in an event, will lead to a better particle identification performance. This is also an aspect we plan to address in the future.

The timing resolution was determined with the setup described in Sec.~\ref{sec:readout} consisting of 1.5-m long strips of EJ-204 scintillator with a rise time of 0.7~ns and S14160-4050HS SiPMs covering 64\% of each end of the strip. We assumed a photodetection efficiency of 50\% based on the datasheet of the SiPM. The effects of the readout electronics have been simulated by using their measured response curve shape.
The timing for each event was then determined from a simple threshold criterion corresponding to the expected signal for one photon. To suppress background, we used only those strips where at least 10 SiPM pixels responded in an event.

\begin{figure}
\centering
\includegraphics[width=0.49\textwidth]{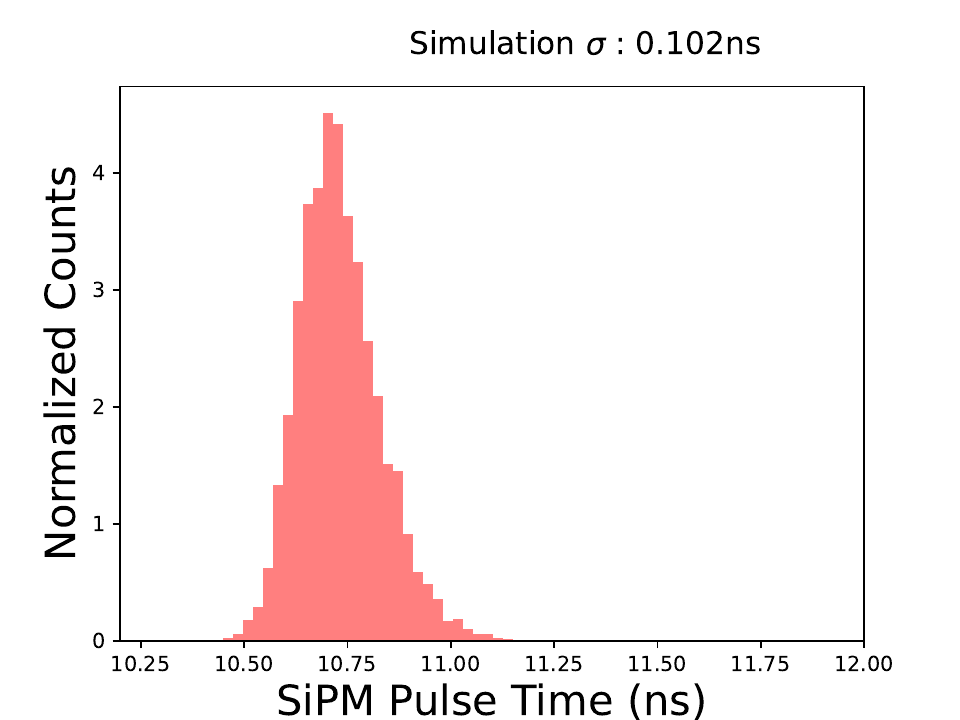}
\caption{Distribution of the arrival time of the first photon using 2~cm-thick scintillator bars including simulated electronics effects.\label{fig:timingDist}}
\end{figure}

The resulting distribution for 2-cm-thick strips with the EJ-204 parametrization is shown in  Fig.~\ref{fig:timingDist}. The resolution is close to the target resolution of 100 ps. For these studies, we assumed tracks going perpendicular through the center of the bar. 
As expected,
we observed that for shorter bars, or equivalently, tracks closer to the readout, the resolution is increasingly 
dominated by the readout electronics. The combined resolution is in line with literature values for similar configurations~\cite{Bohm:2016jhc,gruber2016barrel, Wang:2025stj}.
Assuming no contribution from the readout, the resolution scales as expected with $\sqrt{d}$, where $d$ is the thickness of the scintillator.

\section{ML/AI assisted Detector Optimization}
\label{sec:optimization}
A novel aspect of the design of the hKLM is the use of Multi-Objective Bayesian Optimization (MOBO) for the design optimization. We use the AID2E framework~\cite{diefenthaler2024ai}. 
Four objectives are at the limit of what has been done with MOBO so far and we  
selected MuID performance and energy resolution for neutrons, in the low energy and high energy ranges defined in Sect.~\ref{sec:performance}, as the objectives to be optimized. 
Compared to previous applications of AID2E for the EIC project detector~\cite{diefenthaler2024ai}, the optimization pipeline described here includes automated retraining of a separate ML method for each energy measurement. As we are optimizing four objectives at once, 
there is not necessarily a configuration which is optimal for all objectives at the same time. However, our approach allows us to find configurations that are optimal in the sense that no objective can be improved without degrading another objective, the so-called Pareto front. 
This allows us to explore tradeoffs between the different objectives in a systematic way. The final design will then have to be chosen based on the specific physics cases to be addressed most optimally.


Automated optimization also offers the opportunity to investigate a more complex parameter space. To illustrate the method, we fix, in addition to the inner radius, the outer radius of the Barrel detector to the nominal size allowed by the current hall setup of $r_{out}=2835$~mm, and investigate the possibility of changing the thickness of the steel and scintillator layers independently and linearly along the radius. This is discussed in Sec.~\ref{sec:moboLinDep}. As this is likely difficult to implement in practice, we also looked at the possibility of having two distinct regions of the detector with different steel/scintillator ratios, akin to a pre-shower. This is discussed in Sec.~\ref{sec:moboPreshower}. 

\subsection{Impact of Steel Ratio}
\begin{figure}
\centering
\includegraphics[width=0.45\textwidth]{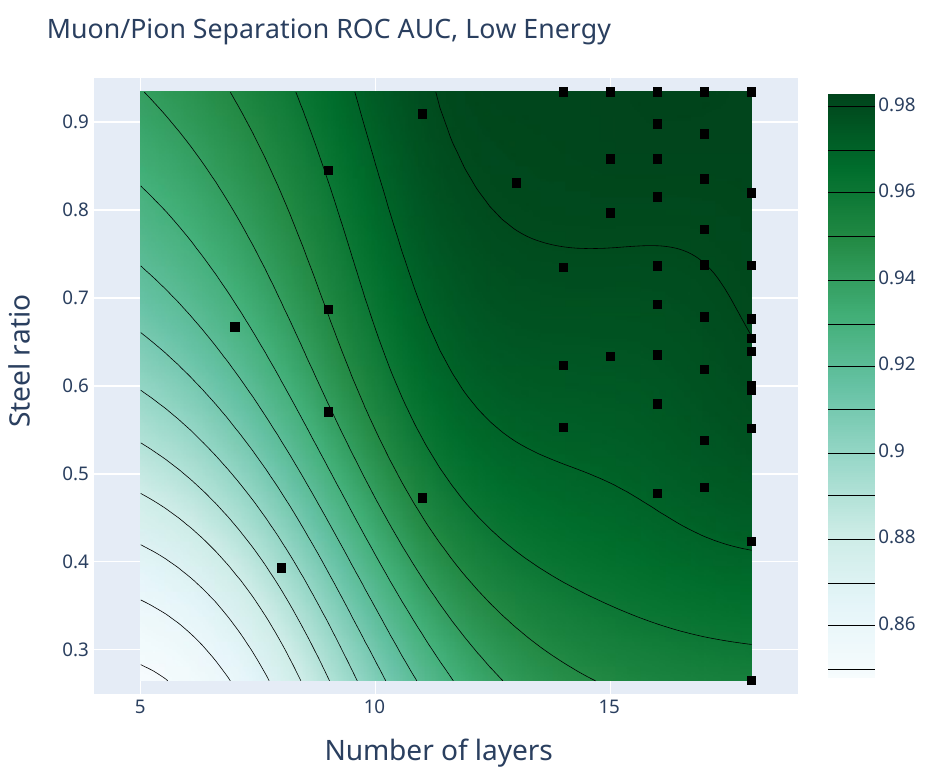}
\includegraphics[width=0.45\textwidth]{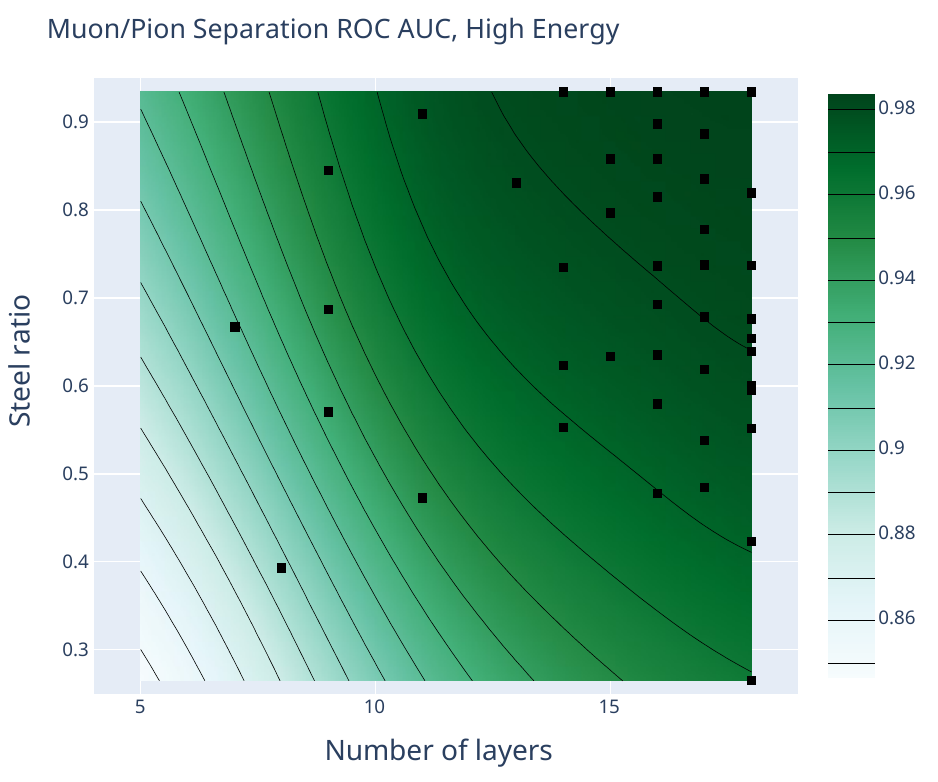}
\includegraphics[width=0.45\textwidth]{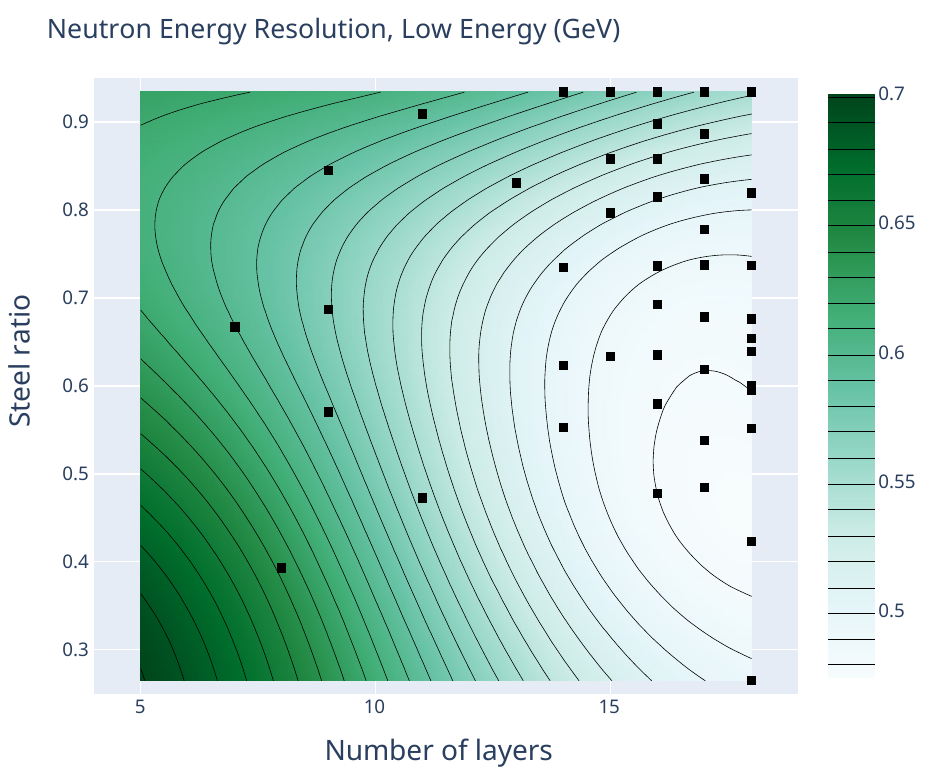}
\includegraphics[width=0.45\textwidth]{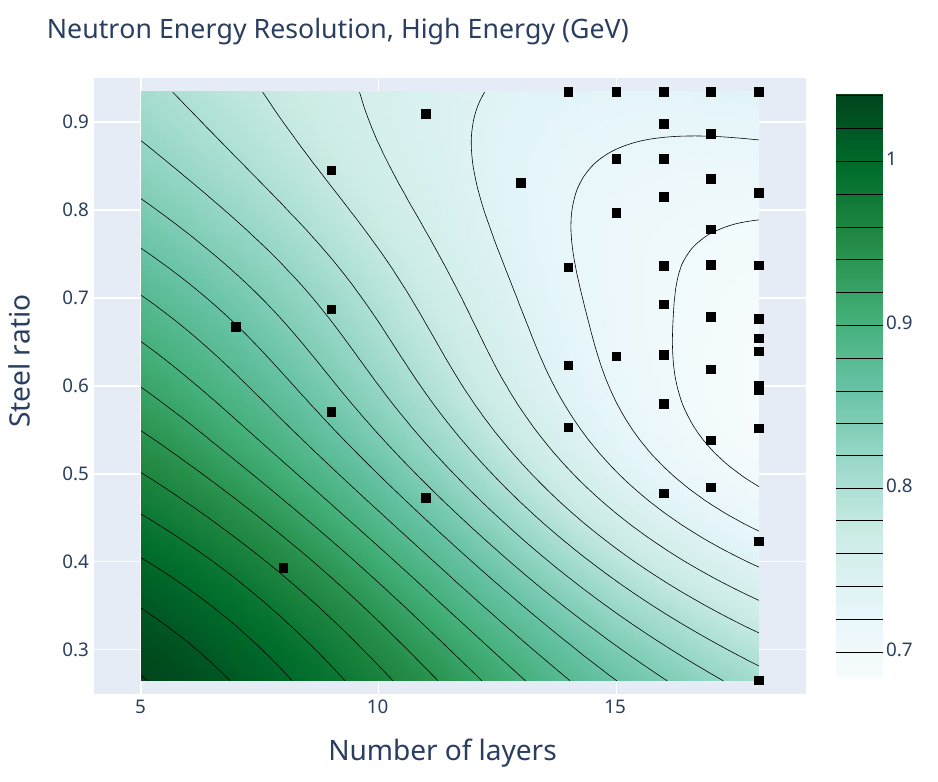}
    \caption{
     Contours of MuID and Energy resolution performance plotted \textit{vs.} the number of layers and the steel ratio. The steel ratio is the ratio of the steel thickness to the sum of the steel and scintillator thicknesses within each layer. Equal color shades represent equal performance in  $\mu/\pi$ separation (upper plots) or energy resolution (RMSE, lower plots). Black points show parameters used in specific trials.
    Upper panel, left to right: $\mu/\pi$ separating power for low energy and high energy momentum ranges; lower panel, left to right: resolution of the energy measurement for neutrons with low energy and high energy momentum ranges.
    See text for detailed discussion.
    \label{fig:moboBasic}}
\end{figure}

\begin{figure}


\includegraphics[width=0.95\textwidth]{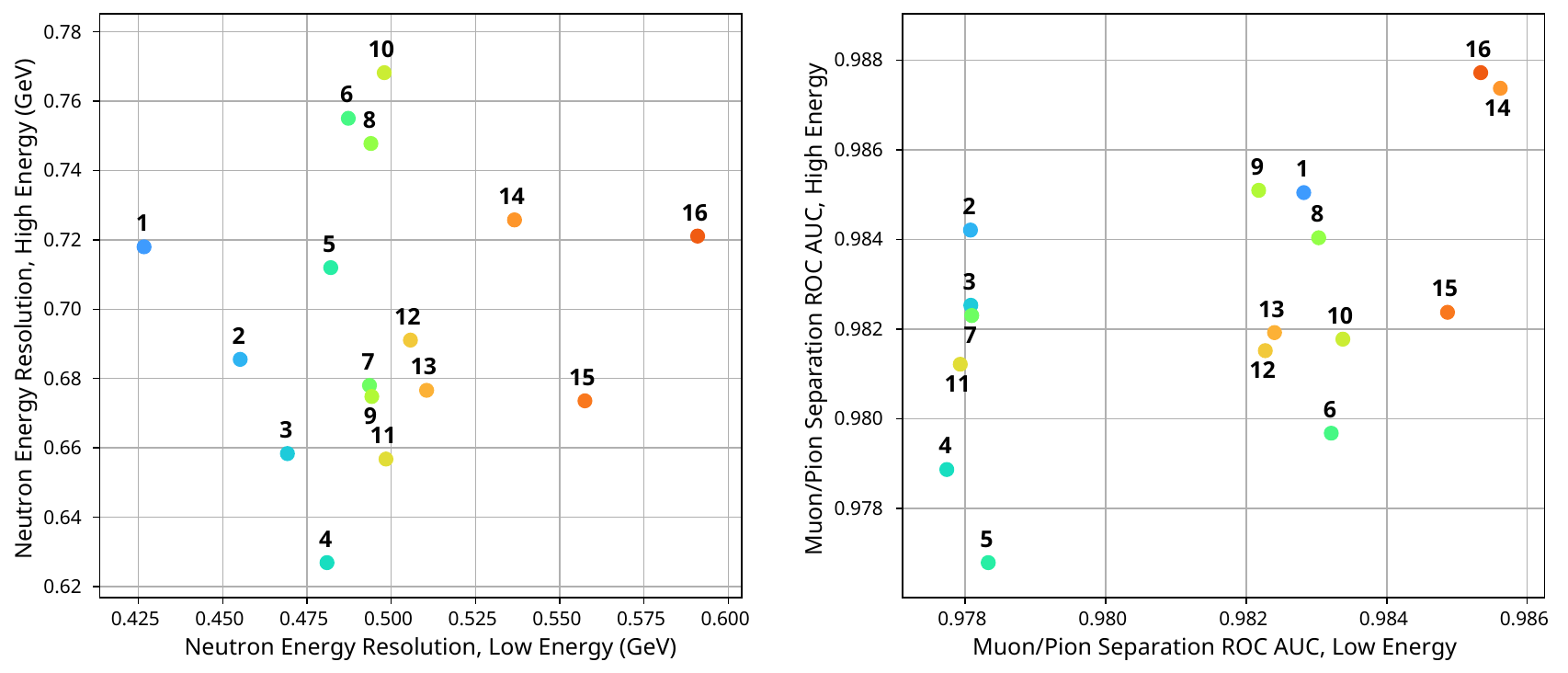}

\caption{Projections of example configurations that lie on the four dimensional Pareto front on the pairwise objectives high and low energy neutron RMSE (left) and MuID at high and low momenta (right). The points in the figures are examples on the Pareto front. Selected points are labeled and correspond to the entries in Table~\ref{tab:trials_basic} and denote the same configurations in both plots. Each point is described by two values, the ratio of steel and the number of layers in the hKLM. See text for more details.
\label{fig:moboBasicPareto}}

\captionof{table}{Geometry parameter values for the trials shown in Figure~\ref{fig:moboBasicPareto}.}
\label{tab:trials_basic}
\resizebox{\textwidth}{!}{%
\begin{tabular}{l|cccccccccccccccc}
\hline
\textbf{Parameter} & \textbf{1} & \textbf{2} & \textbf{3} & \textbf{4} & \textbf{5} & \textbf{6} & \textbf{7} & \textbf{8} & \textbf{9} & \textbf{10} & \textbf{11} & \textbf{12} & \textbf{13} & \textbf{14} & \textbf{15} & \textbf{16} \\
\hline
\texttt{Number of layers}  & 17 & 17 & 16 & 18 & 18 & 17 & 17 & 16 & 18 & 14 & 18 & 15 & 18 & 16 & 15 & 17 \\
\texttt{Steel ratio} & 0.68 & 0.48 & 0.74 & 0.55 & 0.60 & 0.84 & 0.78 & 0.82 & 0.82 & 0.82 & 0.60 & 0.80 & 0.74 & 0.90 & 0.86 & 0.93 \\
\hline
\end{tabular}%
}
\end{figure}

A first simple and exploratory study is to change the ratio between steel and scintillator in the conventional design described above. Here we refer to the ratio of the steel thickness to the total thickness 
($\mathrm{steel~ thick.} / (\mathrm{steel~ thick.} + \mathrm{scint.~ thick.})$ as the steel ratio. Figure~\ref{fig:moboBasic} shows the results for this study for each objective and Fig.~\ref{fig:moboBasicPareto} the resulting examples on the Pareto front. Table~\ref{tab:trials_basic} lists the parameters corresponding to each result 
plotted in Figure~\ref{fig:moboBasicPareto}. We note that the full Pareto front has the dimensionality of the objective space (so, four in this case for two momentum ranges in MuID performance and two momentum ranges for HCAL performance) and is therefore difficult to visualize. 
Here we chose to show projections of exemplary configurations that lie on the Pareto front on two different pairs of objectives with consistent labeling (high energy $vs.$ low energy calometric performance and high-energy $vs.$ low-energy MuID performance). Therefore, when interpreting, \textit{e.g.}, Fig.~\ref{fig:moboBasicPareto}, it is important to note that the two subfigures shown on the left and right are correlated. While the configurations that optimize the calorimetric performance lie in the lower left of the left subfigure, they might not optimize the MuID performance (right subfigure); the MuID performance is in general optimized for points in the upper right corner of the right subfigure. Since all displayed points lie on the Pareto frontier, there does not exist a solution that improves the performance for any one of the objectives without degrading the performance for another objective. The status quo configuration described in Sec.~\ref{sec:Design} does not lie on the Pareto front here 
in Fig.~\ref{fig:moboBasicPareto}
because there are other configurations that perform better on all four objectives, for example trials 1 and 9. Figure~\ref{fig:moboBasicPareto} shows that with the GNN MuID algorithm, there is little variation in the MuID performance over the parateto front, i.e., an optimal performance for the energy reconstruction does not negatively impact MuID performance. This is different from what we observed with the conventional MuID algorithm. There is more performance variation of the energy reconstruction on the pareto front reflecting similar trends that are already visible in the contour plots. For example, a higher steel ratio seems to degrade neutron energy resolution at low energies, e.g., parameter sets '1' and '2' vs. '15' and '16'.  

As can be seen in Figs.~\ref{fig:moboBasic} and~\ref{fig:moboBasicPareto}, the performance is mainly dependent on the steel-to-scintillator ratio. 
 The steel ratio depends on the energy regime, with lower energies around 1 GeV/$c$ preferring a ratio closer to 50\%. Here, the higher fraction of active scintillator material is beneficial for the energy measurements.
For higher momenta, the preferred iron fraction rises to about 80\% for 5~GeV/$c$ particles
due to the limited detector radius and punch through (muon) and energy leakage (netural hadron calorimetry) effects. 
A larger number of layers is in general beneficial. This is most pronounced at lower energies and for the energy resolution measurements. As shown in the contour plot of the energy resolution performance for high momentum particles in Fig.~\ref{fig:moboBasic} (bottom right), a higher number of layers can compensate for a lower amount of steel. Here, contour lines of equal performance run from points with a high ratio of steel and a modest number of layers to a lower amount of steel and more layers. For example, the same performance with 80\% steel and 10 layers can be achieved with about 50\% steel and 15 (thinner) layers.
This may indicate that the ML based reconstruction can make use of the shower shape.  As expected, Muon identification becomes more difficult for a limited number of hadronic interaction lengths and the dependence of the $\mu-\pi$ separation power on the steel/scintillator ratio is more pronounced. 

\subsection{Steel/Scintillator Ratio Dependence on Radius}
\label{sec:moboLinDep}

\begin{figure}
\centering

\includegraphics[width=0.95\textwidth]{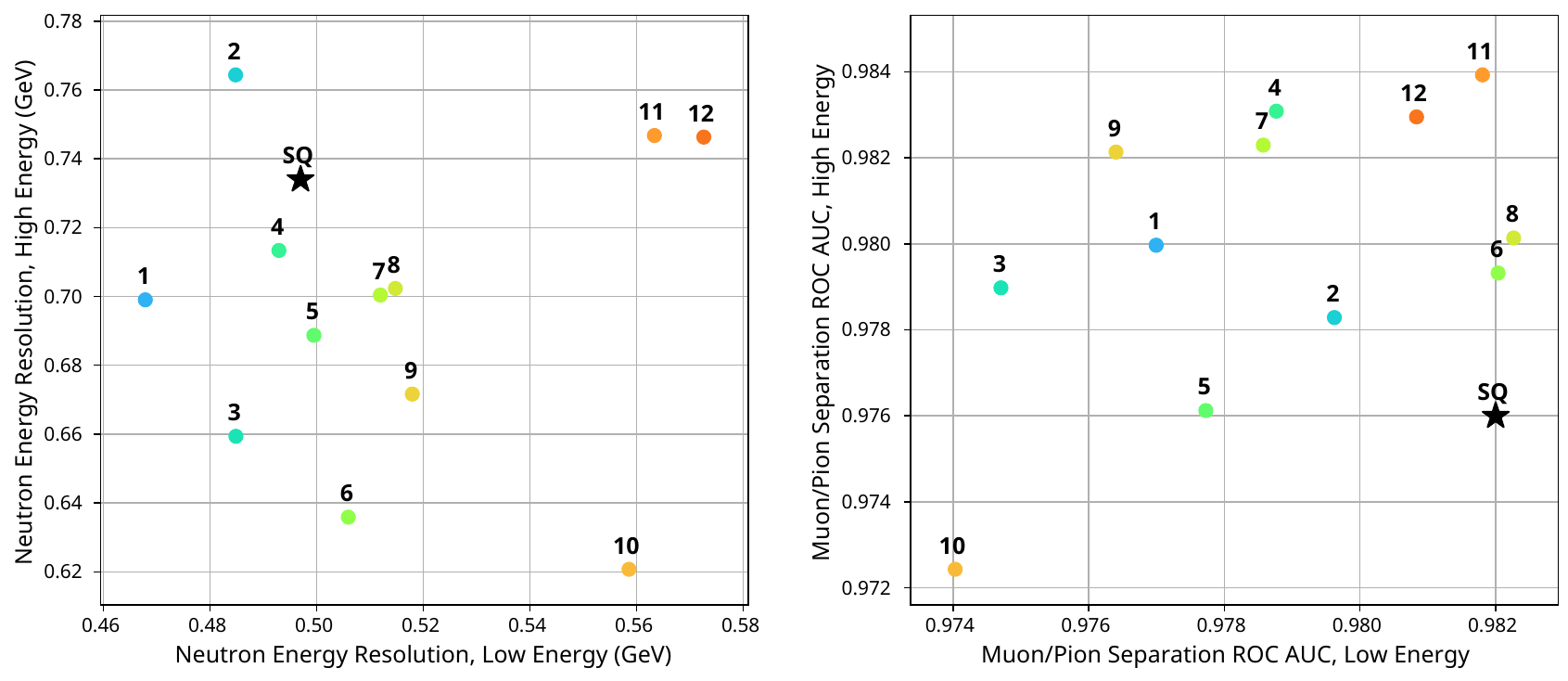}

\caption{Projections of example configurations that lie on the four dimensional Pareto front on the pairwise objectives high and low energy RMSE (left) and MuID at high and low momenta (right). The points in the figures are examples on the Pareto front. Selected points are labeled corresponding to the entries in Table~\ref{tab:trials_linear} and denote the same configurations in both plots. Each point is described by three parameters, the ratio of steel in the hKLM, the slope (rate of change) of increase in the scintillator thickness, and the slope of increase for the steel thickness.
A negative slope indicates that the layers get thinner at the outer radius. A slope of $+1\textrm{ } (-1)$ indicates that the layer thickness goes to 0 at the inner (outer) radius. 
The point labeled ``SQ'' designates the status quo design with fixed scintillator and steel layer thicknesses. See text for more details.
\label{fig:paretoLinear}}

\captionof{table}{Geometry parameter values for the trials shown in Figure~\ref{fig:paretoLinear}.}
\label{tab:trials_linear}
\resizebox{\textwidth}{!}{%
\begin{tabular}{l|ccccccccccccc}
\hline
\textbf{Parameter} & \textbf{1} & \textbf{2} & \textbf{3} & \textbf{4} & \textbf{SQ} & \textbf{5} & \textbf{6} & \textbf{7} & \textbf{8} & \textbf{9} & \textbf{10} & \textbf{11} & \textbf{12} \\
\hline
\texttt{Steel slope}  & 0.80 & 0.54 & 0.43 & 0.50 & 0.00 & 0.56 & 0.80 & 0.66 & 0.80 & 0.32 & 0.80 & $-$0.12 & 0.12 \\
\texttt{Scint. slope}  & 0.29 & 0.80 & 0.15 & 0.26 & 0.00 & 0.42 & 0.80 & 0.80 & 0.40 & 0.07 & 0.56 & $-$0.09 & 0.67 \\
\texttt{Steel ratio}  & 0.90 & 0.90 & 0.73 & 0.90 & 0.72 & 0.88 & 0.90 & 0.85 & 0.79 & 0.89 & 0.70 & 0.74 & 0.85 \\
\hline
\end{tabular}%
}
\end{figure}

A straightforward extension to changing the ratio between steel and scintillator uniformly is to make this ratio dependent on the radius. For simplicity, we considered a linear model in which the slope of change for both is optimized in addition to the overall ratio between scintillator and steel. This gives a total of three parameters: the two slopes and the steel ratio in the detector. Example results for the Pareto front of pairwise objectives in these studies are shown in Fig.~\ref{fig:paretoLinear}. The parameters for each plotted result are listed in Table~\ref{tab:trials_linear}. As can be seen in the figure, muon separation and energy resolution at low and high momentum prefer thicker scintillators towards the outer radius and MuID prefers a larger slope. Compared to the status quo design, a slight improvement for MuID at high momenta can be reached. For the energy measurement, a more substantial improvement of about 15\% can be reached for high energy neutrons. As in the case for a uniform steel-to-scintillator ratio, a higher steel-to-scintillator ratio is helpful at larger particle momenta. It is non-trivial to derive a general pattern from Fig.~\ref{fig:paretoLinear}, as performance depends on the parametrized topology of the detector in relation to the spatial development of the signal. However, some general observations can be made. As in the more simple configuration discussed previously, there is little variation in the MuID results. For neutron energy resolution at lower energies, a high steel ratio coupled with relatively high scintillator and iron slopes seems to be beneficial. At high energies, performance benefits from higher steel slope values that are larger or at least equal to the scintillator slope values. An example is the increase in performance going from parameter sets '10' or '6' to sets '2', '11' or'12'.

In contrast to Fig.~\ref{fig:moboBasicPareto}, the status quo configuration lies on the Pareto front for this 
study,
as shown in Fig.~\ref{fig:paretoLinear}. The inclusion of the status quo suggests that no trial configuration simultaneously outperforms the status quo configuration on every objective. This can be seen by noticing that trials 1, 2, 3, 4, 5, 7, and 9 lie strictly to the left and below the status quo in the left panel of Fig.~\ref{fig:paretoLinear}, meaning they perform better on both neutral hadron energy reconstruction objectives. However, none of these points are strictly to the right and above the status quo in the right panel, indicating that the status quo outperforms each trial on some objective. The same exercise can be done with Fig.~\ref{fig:paretoLayerSplit} to see why the status quo lies on the Pareto front for this third 
study.

\subsection{Adding Pre-shower Layers}
\label{sec:moboPreshower}
\begin{figure}
\includegraphics[width=0.49\textwidth]{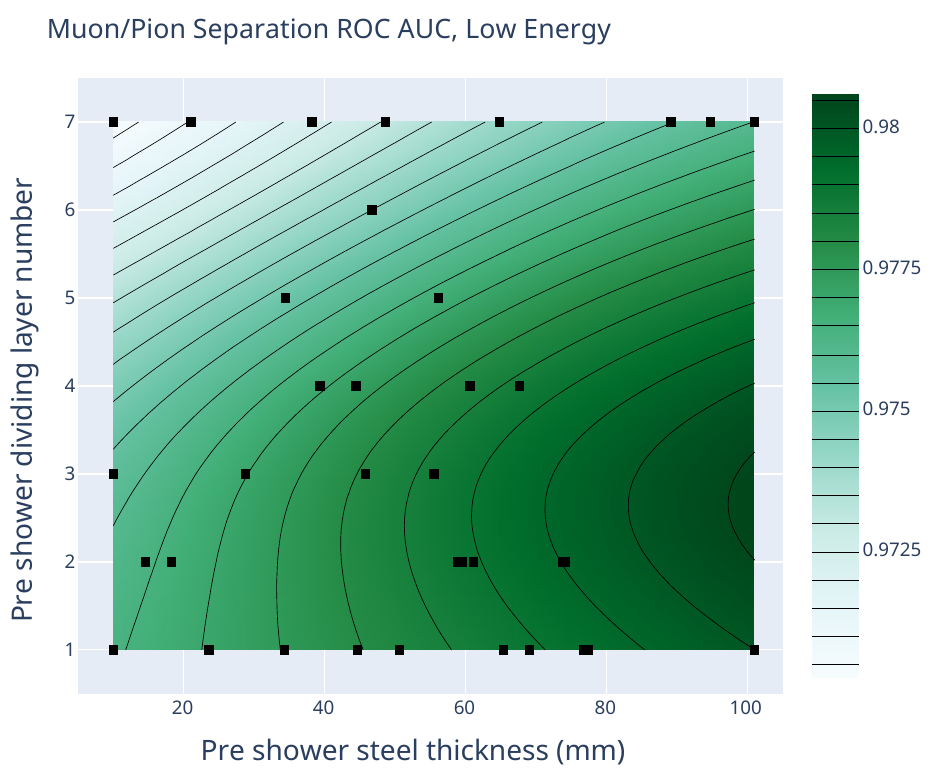}
\includegraphics[width=0.49\textwidth]{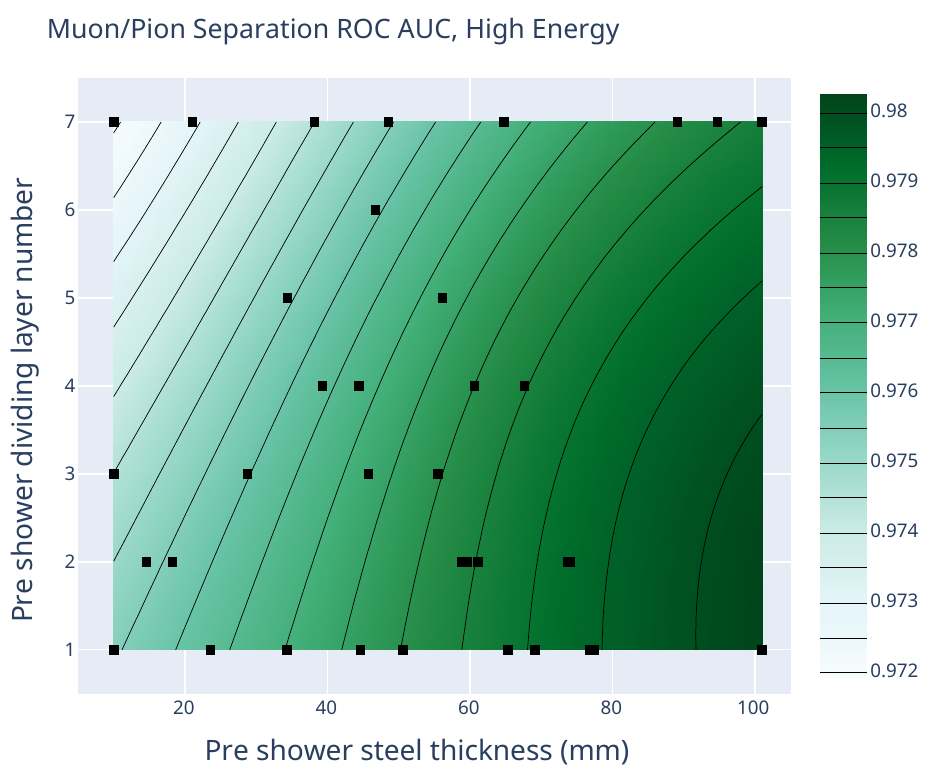}\\
\includegraphics[width=0.49\textwidth]{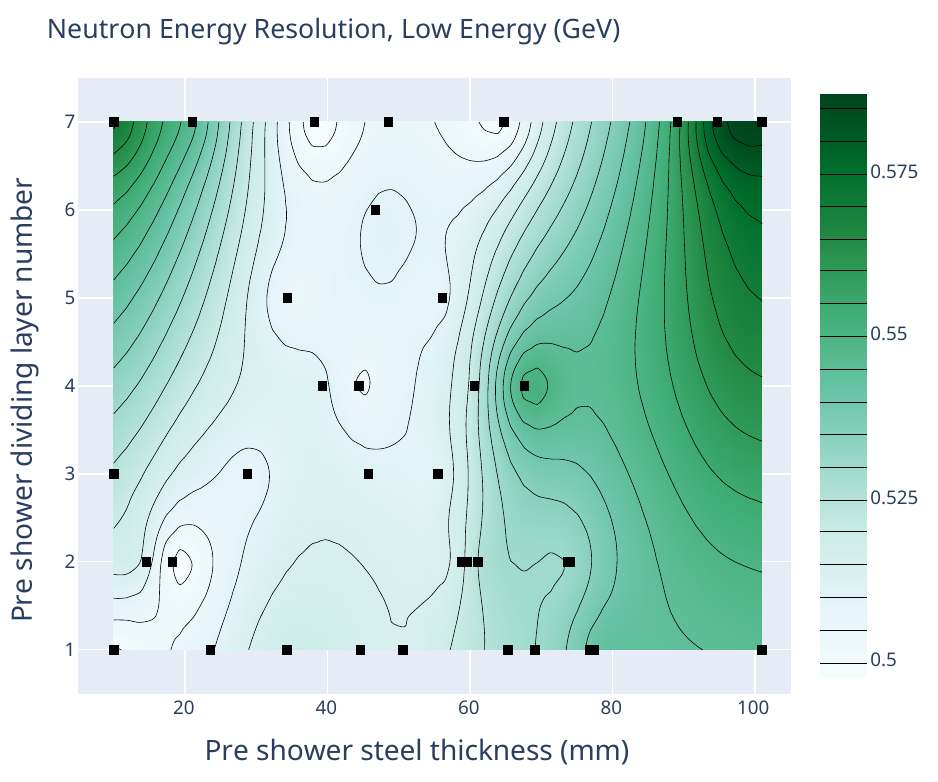}
\includegraphics[width=0.49\textwidth]{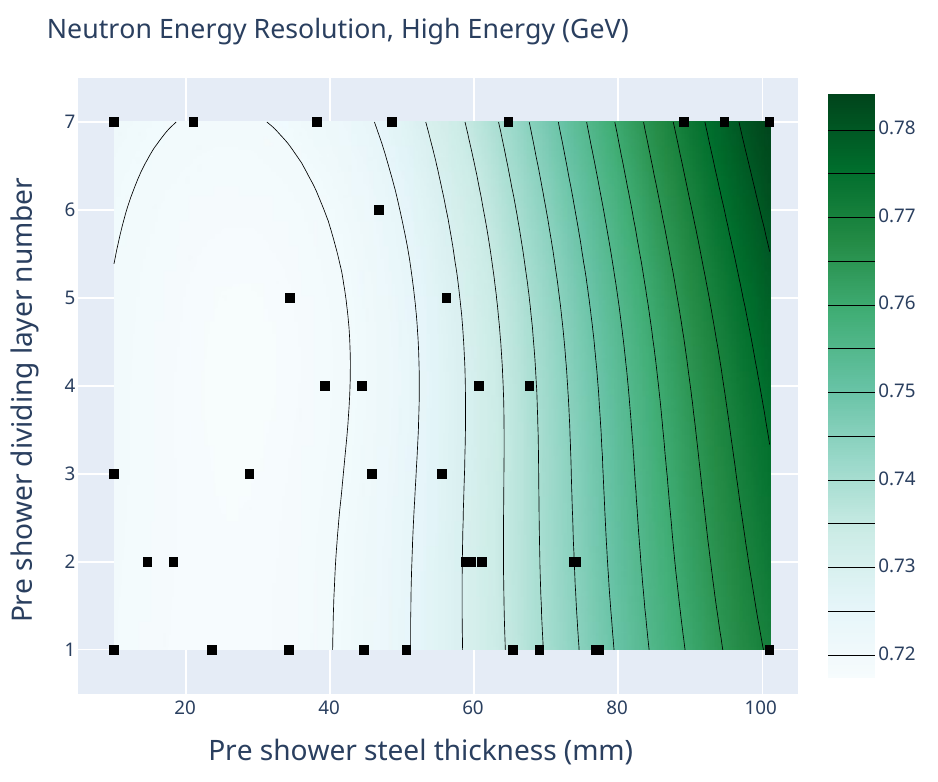}
\caption{Contours of MuID and Energy resolution performance plotted \textit{vs.} the dividing layer number and the pre-shower steel thickness. Equal color shades represent equal performance in  $\mu/\pi$ separation (upper plots) or energy resolution (RMSE, lower plots). Upper panel, left to right: $\mu/\pi$ separating power for low energy and high energy momentum ranges; lower panel, left to right:
energy resolution measurement of neutrons in low energy and high energy momentum ranges.
As discussed in the text, a thickness below 56~mm indicates thinner steel layers in the pre-shower compared to the rest of the detector. 
\label{fig:contTwoParms}}
\end{figure}

Building on the ideas of varying layer thicknesses with radius, and that the different parts of the hKLM (functioning as an HCAL) will provide information for different processes and energy regimes (and hence might profit from a different layout), we also investigated an optimization where we split the standard 14 layer calorimeter along the radial direction and allowed a different steel thickness for each section. Keeping the scintillator thickness fixed at 2~cm, this reduces the optimization problem to two parameters: the layer number where the detector is split and the thickness of the steel in one section. Due to the radial constraint, the latter fixes the steel thickness in the other section. The dependence of the objectives on this two-dimensional parameter space can be visualized in contour plots as shown in Fig.~\ref{fig:contTwoParms}. At high and low momenta, MuID prefers few (2-3) pre-shower layers that are thicker. For the energy resolution at low momenta, a preshower layer with a thickness around 40~mm, slightly lower than the nominal 56~mm, seems best. At high momenta, this should be further reduced to a thickness of about 30~mm.
For both energy ranges the dependence on the dividing layer is quite weak.

\begin{figure}
\centering

\includegraphics[width=0.95\textwidth]{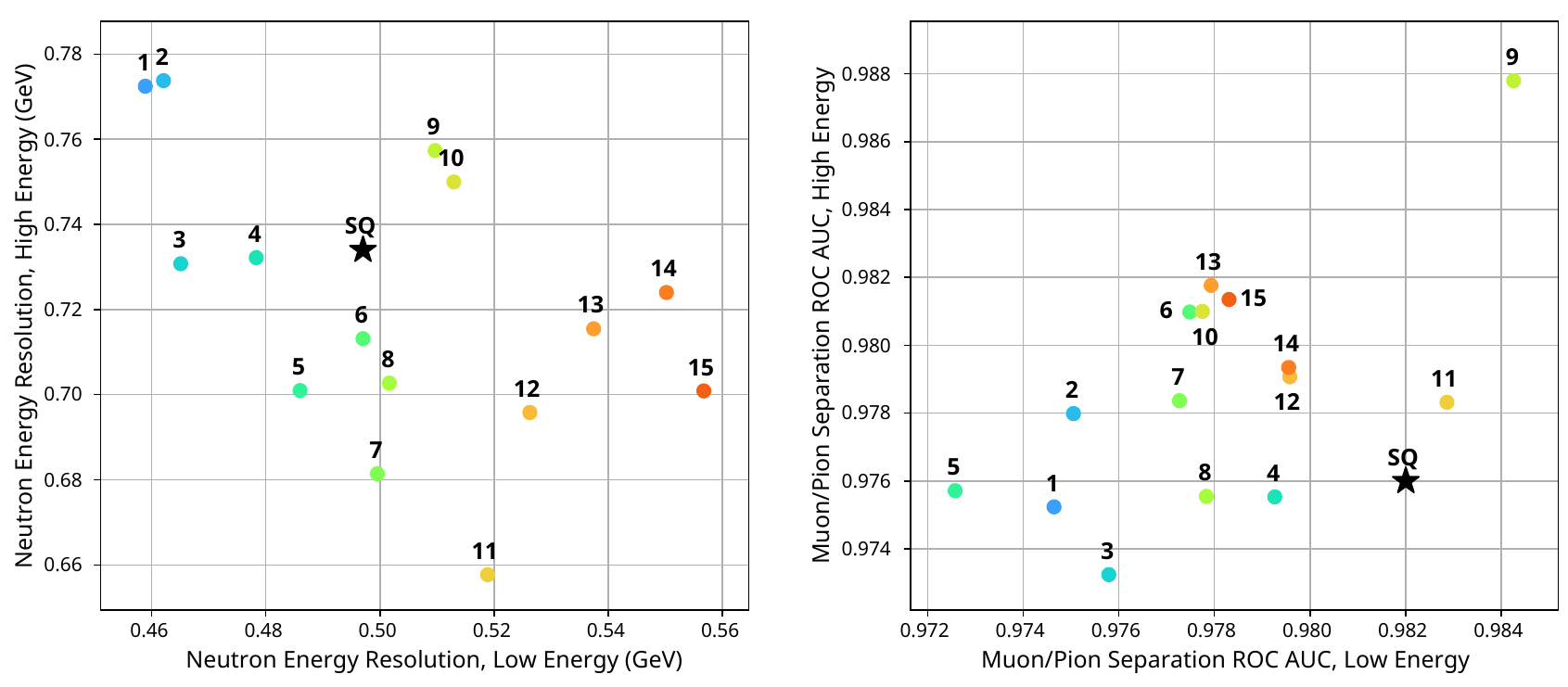}


\caption{Projections of example configurations that lie on the four dimensional Pareto front for the pairwise objectives high and low energy RMSE (left) and MuID at high and low momenta (right). The labels on the examples correspond to Table~\ref{tab:trials_preshower} and indicate the dividing layer number, which is the last one in the pre-shower, and the steel thickness (in mm) for the pre-shower. The label corresponds to the same configuration in both plots. Values below 56~mm indicate thinner steel in the pre-shower than in the rest of the detector and vice versa. The point labeled ``SQ'' designates the status quo design with fixed scintillator and steel layer thicknesses. See text for additional discussion.\label{fig:paretoLayerSplit}}

\captionof{table}{Detector geometry parameters  for the trials shown in Figure~\ref{fig:paretoLayerSplit}.}
\label{tab:trials_preshower}
\resizebox{\textwidth}{!}{%
\begin{tabular}{l|cccccccccccccccc}
\hline
\textbf{Parameter} & \textbf{1} & \textbf{2} & \textbf{3} & \textbf{4} & \textbf{5} & \textbf{SQ} & \textbf{6} & \textbf{7} & \textbf{8} & \textbf{9} & \textbf{10} & \textbf{11} & \textbf{12} & \textbf{13} & \textbf{14} & \textbf{15} \\
\hline
\texttt{Pre shower steel thickness}  & 18.2 & 64.8 & 38.2 & 10.0 & 44.4 & 55.5 & 56.2 & 58.8 & 28.8 & 73.8 & 50.6 & 60.7 & 39.3 & 51.0 & 76.8 & 78.9 \\
\texttt{Dividing layer number} & 2      & 7      & 7      & 1      & 4      & 3      & 5      & 2      & 3      & 2      & 1      & 4      & 4      & 4      & 1      & 3      \\
\hline
\end{tabular}%
}
\end{figure}

Figure~\ref{fig:paretoLayerSplit} shows examples of points on the corresponding Pareto front, and the corresponding parameters are listed in Table~\ref{tab:trials_preshower}. They are labeled by the dividing layer and the steel thickness in mm for the pre-shower. A thickness of 56~mm would indicate the same steel-to-scintllator ratio in the pre-shower as in the rest of the detector. Below 56~mm, the pre-shower has thinner steel layers, and beyond, thicker steel. It appears that the MuID has a slight preference for thicker steel layers in the inner segment and thinner layers at outer radii, whereas energy measurements are better in the reverse setup. At the same time, MuID at lower momenta prefers fewer pre-shower layers than at high momenta. Again, a modest increase of the number of layers does not lead to any improvement when compared to a uniform distribution of steel thickness.

\section{Summary}
We have explored the design of a MuID and hadronic barrel calorimeter based on a scintillator-iron sandwich design, where the iron also serves as the flux return of the central magnet. We show that a multi-dimensionally segmented readout, using machine learning in its design coupled with the resulting analysis capabilities, has the potential to deliver excellent performance for both MuID and hadronic calorimetry. We explored multi-dimensional objective spaces and studied the tradeoffs between the different design parameters. A final choice will depend on the desired performance of the detector in its interplay with other subsystems, which will ideally be co-designed with the hKLM. 
Using a direct readout of the photons from hits in the active-element scintillator strips, a timing resolution of the order of 100~ps in simulation is reached for time-of-flight measurements. This performance also allows for a more compact design using strips running along only one direction per readout superlayer. Compared to other highly segmented calorimeters using an iron-scintillator design, such as the CALICE AHCAL~\cite{CALICE:2024jke}, the moderate multiplicities at the EIC allow us to use scintillator strips instead of tiles. The reduced complexity facilitates a design that can reach the required timing resolution while still maintaining a high granularity.
Furthermore, we have explored several variations of the design in which the steel-to-scintillator ratio is changed along the radial direction. These designs are systematically explored using a MOBO framework from the AID2E project. To enable automatic design optimization, we sped up the optical photon simulation using normalizing flow techniques.
Future studies will include the optimization of the neutral particle identification performance of such a subsystem, potentially taking into account the information available from an EIC detector's other subsystems for the whole event.
\section{Acknowledgments}
This work was funded in part by the generic EIC R\&D program funded under U.S. DOE Field Work Proposal JLAB-NP-13, ``Generic EIC-related Detector Research and Development Program Description Under the Auspices of the Electron-Ion Collider Program Managed by Thomas Jefferson National Accelerator Facility'', Proposal EICGENR\&D2023\_18. The authors acknowledge the support of the U.S. Department
of Energy, Office of Science, Office of Nuclear Physics under contracts DE-SC0024505, DE-SC0024478, and DE-SC0024478 and the U.S. National Science Foundation under awards PHY-2209481, PHY-2514907, and PHY-2412777. 




\section{Bibliography}
\bibliographystyle{elsarticle-num}
\bibliography{biblio}
\end{document}